\documentclass[prd,nofootinbib]{revtex4-1}
\usepackage[T1]{fontenc}
\usepackage[latin9]{inputenc}
\usepackage{amsmath}
\usepackage{amssymb}
\usepackage{graphicx,epsfig}
\begin{document}
\begin{flushleft}
KCL-PH-TH/2012-23 \\
LCTS/2012-28
\end{flushleft}

\title{CPT-Violating Leptogenesis induced by Gravitational Defects}

\vspace{1cm}
\author{Nick E. Mavromatos\footnote{Currently also at: Theory Division, Physics Department, CERN, CH 1211 Geneva 23, Switzerland.} and Sarben Sarkar}

\vspace{1cm}

\affiliation{Theoretical Particle Physics and Cosmology Group, Department of Physics, King's College London,  Strand, London WC2R 2LS, UK.}

\begin{abstract}
We explore leptogenesis induced by the propagation of neutrinos in gravitational backgrounds that may occur in string theory.  The first background is due to
linear  dilatons and the associated Kalb-Ramond field (axion) in four non-compact space-time dimensions of the string, and can be described within the framework of local effective lagrangians. The axion is linear in the time coordinate of the Einstein frame and gives rise to a constant torsion which couples  to the fermion
spin through a gravitational covariant derivative. This leads to different energy-momentum dispersion relations for fermions and anti-fermions and hence leptogenesis.
The next two backgrounds go beyond the \emph{local} effective lagrangian framework. One is a stochastic (Lorentz Violating) Finsler metric which again leads to different dispersion relations between fermions and antifermions.
The third background of primary interest is the one
due to populations of stochastically fluctuating point-like space-time defects that can be encountered in string/brane theory
(D0 branes). Only neutral matter interacts non-trivially with these defects, as a consequence of charge conservation. Hence, such a background singles out neutrinos among the Standard Model excitations as the ones interacting predominantly with the defects. The back-reaction of the defects on the space-time due to their interaction with neutral matter results in stochastic Finsler-like metrics (similar to our second background).
On average, the stochastic fluctuations of the D0 brane defects preserve Lorentz symmetry, but
their variance is                                                                                                                                                                                                                                                                                                                                                                       non-zero. Interestingly, the particle/antiparticle asymmetry comes out naturally to favour matter over antimatter in this third background, once the effects of the kinematics of the scattering of the D branes with matter is incorporated.

\end{abstract}

 \maketitle

\section{Introduction}

One of the most important issues of fundamental physics, relates to
an understanding of the magnitude of the observed baryon asymmetry $n_{B}-n_{\overline{B}}$ (where $B$ denotes baryon, $\overline{B}$ denotes antibaryon, $n_{B}$ is the number density of baryons and $n_{\overline{B}}$ the number density of antibaryons
in the universe). The numbers of protons and neutrons far exceeds the number of antiprotons and antineutrons.The universe is overwhelmingly made up of matter rather than anti-matter. According to the standard Big Bang theory, matter and antimatter have been created in equal amounts in the early
universe. However, the observed charge-parity (CP) violation in particle
physics~\cite{Christenson:1964fg}, prompted A. Sakharov~\cite{Sakharov:1967dj}
to conjecture that for Baryon Asymmetry in the universe (BAU) we need:
\begin{itemize}
\item
Baryon number violation to allow  for states with $\Delta B\neq 0$ starting from states with $\Delta B =0$ where $\Delta B$ is the change in baryon number.
\item
If C or CP conjugate processes to a scattering process were allowed with the same amplitude  then baryon asymmetry would disappear. Hence C and CP need to be broken.
\item
Chemical equilibrium does not permit asymmetries. Hence Sakharov required that chemical equilibrium does not hold during an epoch in the early universe.
\end{itemize}
Hence non-equilibrium physics in the early universe together with
baryon number (B), charge (C) and charge-parity (CP) violating interactions/decays of anti-particles in the
early universe, may result in the observed BAU. In fact there are two types of non-equilibrium processes
in the early universe that can  produce this asymmetry: the first
type concerns processes generating asymmetries between leptons and
antileptons (\emph{leptogenesis}), while the second produces asymmetries
between baryons and antibaryons (\emph{baryogenesis}). The near complete
observed asymmetry today, is estimated in the Big-Bang theory~\cite{Gamow:1946eb}
to imply:
\begin{equation}
\Delta n(T\sim 1~{\rm GeV})=\frac{n_{B}-n_{\overline{B}}}{n_{B}+n_{\overline{B}}}\sim\frac{n_{B}-n_{\overline{B}}}{s}=(8.4-8.9)\times10^{-11}\label{basym}
\end{equation}
at the early stages of the expansion, e.g. for times $t<10^{-6}$~s
and temperatures $T>1$~GeV. In the above formula $n_{B}$ ($n_{\overline{B}}$)
denotes the (anti) baryon density in the universe, and $s$ is the
entropy density. Unfortunately, the observed CP violation within the
Standard Model (SM) of particle physics (found to be of order $\epsilon=O(10^{-3})$
in the neutral Kaon experiments~\cite{Christenson:1964fg}) induces
an asymmetry much less than that in (\ref{basym})~\cite{Kuzmin:1985mm}.
There are several ideas that go beyond the SM (\emph{e.g}.  grand unified
theories, supersymmetry, extra dimensional models \emph{etc.}) which involve
the decays of right handed sterile neutrinos. For relevant important
works on this see \cite{Shaposhnikov:2009zz,Shaposhnikov:2006xi,Lindner:2010wr,Kusenko:2010ik,
Randall:1999ee,Merle:2011yv,Barry:2011wb}.
These ideas lead to extra sources for CP violation that could generate
the observed BAU. Fine tuning and somewhat \emph{ad hoc} assumptions are involved in such scenarios
and the quest for an understanding of the observed BAU still needs
further investigation.

The requirement of non-equilibrium is on less firm ground than the other two requirements of Sakharov,
\emph{e.g.} if the non-equilibrium epoch occurred prior to inflation then its effects would be hugely diluted by inflation.
A basic assumption in the scenario of Sakharov is that \emph{CPT 
symmetry}~\cite{cpttheorem}  (where $T$ denotes time reversal operation) holds in the very early universe which leads to the production
of matter and antimatter in equal amounts. Such \emph{CPT invariance}
is a cornerstone of all known \emph{local} effective \emph{relativistic}
field theories without gravity, which current particle-physics phenomenology
is based upon. It should be noted that the necessity
of non-equilibrium processes in CPT invariant theories can be dropped if the
requirement of CPT is relaxed. This violation of CPT (denoted by CPTV) is the result of a breakdown of Lorentz symmetry (which might happen at ultrahigh energies \cite{Mavromatos:2010pk}). For many
models with CPTV, in the time line of the expanding universe, CPTV  generates
first lepton asymmetries (\textit{leptogenesis}); subsequently through
sphaleron processes or Baryon-Lepton (B-L) number conserving processes
in grand unified theories, the lepton asymmetry can be communicated
to the baryon sector to produce the observed BAU.

Thus, CPTV in the early universe may also obviate the need for including extra sources of CP violation, such as sterile neutrinos and/or supersymmetry, in order to obtain the observed BAU. In this article we will consider CPTV leptogenesis from a non-Riemannian point of view, inspired by a stringy model of gravitational defects and backgrounds interacting with neutral fermions. The explicit non-Riemannian structure that we investigate is Finsler geometry \cite{bao2000introduction} where
momentum as well as position are explicitly involved in its metric. The defects that we will consider
are point-like solitonic structures in some string
theories; they are known as D0-branes (or \emph{D-particles}) \cite{johnson2002d}. Our generic model
involves effectively three-space-dimensional brane uuniverseniverses, obtained from compactification of higher-dimensional branes, which are embedded in a  bulk space punctured by
D-particles~(see e.g. \cite{Ellis:2004ay}). One of these brane universes constitutes our observable world, which moves in the bulk. As a consequence of this motion, D-particles cross the brane.
Open strings for electrically
neutral particles on the brane can attach an end to the D-particle; subsequently this string can detach ~\cite{Douglas:1996yp,Ellis:2005ib}.
Scattering off a \emph{population} of D-particles (D-foam) affects the  kinematics of the stringy matter and leaves an
imprint on the background geometry \cite{Mavromatos:2004bx} on the brane world. This geometry is similar to Finsler metrics but
with stochastic parameters~\cite{Mavromatos:2005bu,Bernabeu:2006av,Mavromatos:2010nk}. We first investigate the consequences for gravitational leptogenesis of this
metric structure in the general setting of Finsler geometry without reference explicitly to D-foam . For the underlying stringy model however, since it is microscopic, we can consider in addition the kinematics of D particle scattering. This kinematical aspect leads naturally to the asymmetry between the particle and anti-particle abundances  having the right sign. The kinematical argument involves recoil kinetic energy does not fit into an effective local field theory approach and represents
a new approach that is relevant to leptogenesis.

The structure of the article will be as follows: in the next
section \ref{sec:cptvmodels} we shall briefly review some
relevant existing models for fermionic asymmetry, which entail CPTV in the early universe as alternatives to scenarios involving sterile neutrinos and/or supersymmetry. 
The discussion will also help to
differentiate between our viewpoint and those prevailing in the literature.
The models for leptogenesis that we will discuss are of gravitational type, namely CPTV-induced differences in the dispersion relations between particles and antiparticles propagating in these backgrounds.
In section \ref{sec:string}, we propose a new model for gravitational leptogenesis which follows broadly an earlier framework \cite{Lambiase:2006md, Debnath:2005wk}, but differs crucially in that the full gravitational multiplet \cite{Zwiebach:789942} that arises in string theory is used. The leptogenesis in this model is due to CPTV dispersion relations between fermions/antifermions, induced by the (constant) \emph{torsion} associated with the antisymmetric Kalb-Ramond tensor field in the gravitational multiplet. The torsion is space-time independent for a particular solution of the conformal invariance equations of the string associated with a linear dilaton. In the proposals discussed in section \ref{sec:cptvmodels}  the space-time independence of the torsion was not guaranteed.
In section \ref{sec:Finsler}, we consider another scenario for gravitational leptogenesis that involves a non-Riemannian  Finsler
metric, with stochastically fluctuating parameters, a variant of which appears in our string/brane model in section \ref{sec:dfoam}. In section \ref{sec:dfoam} we present our model for the stringy universe and the induced CPTV and leptogenesis/baryogenesis.
The basic properties of the model are reviewed briefly in subsection \ref{sec:dfoamprop}, where our observable universe is represented as a
brane world,  punctured by fluctuating ``point-like'' brane Defects (\emph{D-particles} or D0-branes);
the back-reaction of D-particles, during their topologically non-trivial  scattering with neutrinos, results in
local metric distortions. These distortions depend, however,  on \emph{both}  the coordinate and momentum variables of the particles, thus inducing a kind of Finsler geometry.
In subsection \ref{sec:kinD} we discuss the kinematics of scattering of the D-particle (D$0-$brane)
and stringy matter in addition to an induced Finsler-like metric. This together with the
stochasticity of the recoil process
leads to CPTV and matter dominance over antimatter \emph{naturally}, without the need for an adjustment of the sign of
the lepton/antilepton asymmetry. This is an interesting feature of our model which differentiates it from
earlier proposals on gravitational leptogenesis/baryogenesis~\cite{Dvali:1999gf, Davoudiasl:2004gf, Li:2004hh, Mukhopadhyay:2005gb, Debnath:2005wk, Lambiase:2006md, Lambiase:2011by, Mukhopadhyay:2007vca, Sinha:2007uh},
where the sign of the asymmetry is implicitly chosen.
In common with  earlier discussions
of CPTV in baryogenesis \cite{Carroll:2005dj}, this class of models
involves violation of Lorentz symmetry, but only because of a
non-zero \emph{variance} of the stochastic parameter. Estimates on the range of parameters of the model, for which the induced asymmetry can lead to
the observed Baryon Asymmetry in the universe, are given. We conclude that D-particles of Planck-size mass  are probably needed for this purpose, if
standard cosmological properties of neutrinos are assumed. Conclusions and
outlook are finally given in section \ref{sec:concl}.

\section{Models of CPT Violation (CPTV) \label{sec:cptvmodels}}

In this section we shall review some existing models of CPTV induced asymmetry between matter and antimatter in the early universe, which can be contrasted with our approach in this article. We shall be brief in our exposition, referring the interested reader to the relevant literature for more details.

\subsection{CPTV Models with Particle-Antiparticle Mass Difference}

The simplest possibility~\cite{Dolgov:2009yk} for inducing CPTV
in the early universe is through particle-antiparticle mass differences
$m\ne\overline{m}$. These would affect the particle phase-space distribution
function $f(E,\mu)$
\begin{equation}
f(E,\mu)=[{\rm exp}(E-\mu)/T)\pm1]^{-1}~,\quad E^{2}=\vec{p}^{2}+m^{2}\label{cptvf}
\end{equation}
and antiparticle phase-space distribution function
\[
f(\overline{E},\bar{\mu})=[{\rm exp}(\bar{E}-\bar{\mu})/T)\pm1]^{-1}~,\quad\bar{E}^{2}=\vec{p}^{2}+\bar{m}^{2},
\]
with $\vec{p}$ being the $3-$momentum. (Our convention will be that
an overline over a quantity will refer to an antiparticle, $+$ will denote a fermionic (anti-)particle and $-$ will denote a bosonic (anti-)particle.) Mass differences
between particles and antiparticles, $\overline{m}-m\neq0$, generate
a matter-antimatter asymmetry in the relevant densities
\begin{equation}
n-\overline{n}=g_{d.o.f.}\int\frac{d^{3}p}{(2\pi)^{3}}[f(E,\mu)-f(\overline{E},\bar{\mu})]~,\label{nnbarcptv}
\end{equation}
 where $g_{d.o.f.}$ denotes the number of degrees of freedom of the
particle species under study. In the case of spontaneous Lorentz violation
\cite{Carroll:2005dj} there is a vector field $A_{\mu}$ with a non-zero
time-like expectation value which couples to a global current $J^{\mu}$
such as baryon number through an interaction lagrangian density
\begin{equation}
\mathcal{L}=\lambda A_{\mu}J^{\mu}.\label{LorVioln}
\end{equation}
This leads to $m\neq\bar{m}$ and $\mu\neq\bar{\mu}$. Alternatively,
following ~ \cite{Dolgov:2009yk} we can make the assumption that
the dominant contributions to baryon asymmetry come from quark-antiquark
mass differences, and that their masses ``run'' with the temperature i.e. 
$m\sim gT$ (with $g$ the QCD coupling constant). One can provide
estimates for the induced baryon asymmetry on noting that the maximum
quark-antiquark mass difference is bounded by the current experimental
bound on the proton-antiproton mass difference, $\delta m_{p}(=|m_{p}-\overline{m}_{p}|)$,
known to be less than $\,2\cdot10^{-9}$ GeV. Taking $n_{\gamma}\sim0.24\, T^{3}$ (the
photon equilibrium density at temperature $T$) we have~\cite{Dolgov:2009yk}:

\begin{equation}
\beta_{T}=\frac{n_{B}}{n_{\gamma}}=8.4\times10^{-3}\,\frac{m_{u}\,\delta m_{u}+15m_{d}\,\delta m_{d}}{T^{2}}~,\quad\delta m_{q}=|m_{q}-{\overline{m}}_{q}|~.
\end{equation}
$\beta_{T}$ is too small compared to the observed one. To reproduce
the observed $\beta_{T=0}\sim6\cdot10^{-10}$ one would need $\delta m_{q}(T=100~{\rm GeV})\sim10^{-5}-10^{-6}~{\rm GeV}\gg\delta m_{p}$,
which is somewhat unnatural.

However, active (\emph{light}) neutrino-antineutrino mass differences
alone may reproduce BAU; some phenomenological models in this direction
have been discussed in \cite{Barenboim:2001ac}, considering, for
instance, particle-antiparticle mass differences for active neutrinos
compatible with current oscillation data. This leads to the result
\begin{equation}
n_{B}=n_{\nu}-n_{{\overline{\nu}}}\simeq\frac{\mu_{\nu}\, T^{2}}{6}
\end{equation}
 yielding $n_{B}/s\sim\frac{\mu_{\nu}}{T}\sim10^{-11}$ at $T\sim100$~GeV,
in agreement with the observed BAU. (Here $s$, $n_{\nu},\,\mathrm{and}\,\mu_{\nu}$
are the entropy density, neutrino density and chemical potential respectively.)

\subsection{CPTV Decoherence Models }

But particle-antiparticle mass differences may not be the only way
by which CPT is violated. As discussed in \cite{Barenboim:2004ev,Barenboim:2004wu},
quantum gravity fluctuations in the structure of space-time, 
may be strong in the early universe; the fluctuations may act as an \emph{environment}
inducing decoherence for the (anti-)neutrinos. However the couplings
between the particles and the environment are different for
the neutrino and antineutrino sectors. Once there is decoherence for
an observer with an energy for a low-energy (compared to the Planck
scale $M_{P}\sim10^{19}$\, GeV) observer, the effective CPT symmetry
generator may be \emph{ill-defined} as a quantum mechanical operator,
according to a theorem by R.~Wald~\cite{Wald:1980nm}, leading to
an intrinsic violation of CPT symmetry. This type of violation may
characterise models of quantum gravity with stochastic space-time
fluctuations due, for instance, to gravitational space-time defects,
as is the case of certain brane models~\cite{Ellis:2005ib,Ellis:2004ay,Bernabeu:2006av}.
In such a case,  a slight mismatch in the strength of
the stochastic space-time fluctuations between particle and antiparticle
sectors, can lead to different decoherence parameters to describe the
interaction of the gravitational environment with matter.

In \cite{Barenboim:2004ev,Barenboim:2004wu}, simple models of Lindblad
decoherence~\cite{Lindblad:1975ef}, conjectured to characterise
quantum-gravity-induced CPTV decoherent situations~\cite{Ellis:1983jz,Ellis:1995xd},
have been considered for neutrinos~\cite{Benatti:2001fa}. It was
assumed on phenomenological grounds, that non-trivial decoherence
parameters were \emph{only} present in the antiparticle sector: this
is consistent with the lack of any experimental evidence to date~\cite{Fogli:2007tx,Lisi:2000zt,Barenboim:2006xt}
for vacuum decoherence in the particle sector. The antineutrino decoherence
parameters (with dimension of energy) had a mixed energy dependence.
A diagonal Lindblad decoherence matrix for three-generation neutrinos
requires eight coefficients $\overline{\gamma}_{i}$. Some of the
eight coefficients were assumed for simplicity in \cite{Barenboim:2004ev,Barenboim:2004wu}
to be proportional to the antineutrino energy
\[
\overline{\gamma}_{i}=\frac{T}{M_{P}}\, E~,\quad i=1,2,4,5
\]
 while the remaining (subdominant) ones were inversely proportional
to it
\[
\overline{\gamma}_{j}=\frac{10^{-24} \, ({\rm GeV})^2}{E}~,~j=3,6,7,8~.
\]
 The model was proposed without any microscopic justification; its
choice was originally motivated by fitting the LSND ``anomalous data''
in the antineutrino sector~\cite{Aguilar:2001ty} with the rest of
the neutrino data. and this required $T$ to be $T/M_{P}\sim10^{-18},$i.e.
in the temperature range of electroweak symmetry breaking. One can
derive~\cite{Barenboim:2004ev,Barenboim:2004wu} an active (light)
$\nu-\overline{\nu}$ asymmetry of order
\begin{equation}
\mathcal{A}=\frac{n_{\nu}-n_{\overline{\nu}}}{n_{\nu}+n_{\overline{\nu}}}=\frac{\overline{\gamma}_{1}}{\sqrt{\Delta m^{2}}}=\frac{T}{M_{P}}\cdot\frac{E}{\sqrt{\Delta m^{2}}}~,\label{nnubar}
\end{equation}
where $\Delta m^{2}$ denotes the (atmospheric) neutrino mass squared
difference, which plays the r�le of a characteristic low mass scale
in the problem. This lepton number violation is communicated to the
baryon sector by means of baryon number ($B$) plus lepton number
($L$) conserving sphaleron processes. These processes lead to an
estimate~\cite{Barenboim:2004ev} for the current value of $B$ to
be
\begin{equation}
B=\frac{n_{\nu}-n_{\overline{\nu}}}{s}\sim\mathcal{A}\,\frac{n_{\nu}}{g^{\star}\, n_{\gamma}}\label{baryon}
\end{equation}
with $n_{\gamma}$ the photon number density, $g^{\star}$ the effective
number of degrees of freedom (at the temperature where the asymmetry
developed, i.e. the electroweak symmetry breaking temperature in the
model of \cite{Barenboim:2004ev}). $g^{\star}$ depends on the matter
content of the model (with a typical range $g^{\star}\in[10^{2}-10^{3}]$).
For such parameter values $\mathcal{A}\sim10^{-6}$ and so the observed
BAU may be reproduced in this case without the need for extra sources
of CP violation e.g. sterile neutrinos.
Such models, however, do not provide an underlying microscopic understanding.
In particular there is missing an understanding of the preferential role of the neutrino
compared to other particles of the Standard Model in the CPT violating
decoherence process. Within some microscopic models of
space-time foam, involving populations of point-like brane defects
(D-particles) puncturing three(or higher)-spatial dimension brane
worlds~\cite{Ellis:2005ib,Ellis:2004ay,Bernabeu:2006av}, such a
preferred role may be justified as we shall discuss in  section
\ref{sec:dfoam}. Moreover D-particles also imply that the framework
of Riemannian geometry will need to be generalised to Finsler geometries.
Finsler geometries (with a stochastic background) will be investigated
in order to evaluate the possibilities of CPTV occuring.

\subsection{CPTV-induced by Curvature effects in Background Geometry}\label{sec:bg}

Although the role of gravity was alluded to in the last subsection,
associated features of space-time were not discussed. In the literature
the role of gravity has been explicitly considered within a local effective
action framework which is essentially that of (\ref{LorVioln}) A
coupling to scalar curvature $\mathcal{R}$ \cite{Davoudiasl:2004gf,Lambiase:2006md,Lambiase:2011by,Li:2004hh} through
a CP violating interaction Lagrangian $\mathcal{L}$:
\begin{equation}
\mathcal{L}=\frac{1}{M_{*}^{2}}\int d^{4}x\sqrt{-g}\left(\partial_{\mu}\mathcal{R}\right)J^{\mu}\label{gravBaryogen}
\end{equation}
where $M_{*}$ is a cut-off in the effective field theory and $J^{\mu}$ could be the current associated with $B-L$ ( $L$ being the lepton number).
There is an implicit choice of sign in front of the interaction (\ref{gravBaryogen}), which has been fixed so as to ensure matter dominance.

It has been shown that \cite{Davoudiasl:2004gf}
\begin{equation}\label{leptoasymm}
    \frac{n_{B-L}}{s}=\frac{\dot{R}}{M_{*}^{2} T_{d}},
\end{equation}
$T_{d}$ being the freeze-out temperature for $B-L$ interactions. The idea then is that this asymmetry can be converted to baryon number asymmetry provided the $B+L$ electroweak sphaleron interaction has not frozen out.
To leading order in $M_{*}^{-2}$ we have $R=8\pi G (1-3w) \rho$ where $\rho$ is the energy density of matter and the equation of state is $p=w\rho$ where $p$ is pressure. For radiation $w=1/3$ and so in the radiation dominated era of the Friedmann-Robertson-Walker cosmology $R=0$. However $w$ is  precisely $1/3$ when $T^{\mu}_{\mu}=0$. In general $T^{\mu}_{\mu}\propto \beta(g)F^{\mu\nu}F_{\mu\nu}$  where $\beta(g)$ is the beta function of the running gauge coupling $g$ in a $SU(N_{c}$ gauge theory with $N_{c}$ colours. This allows $w\neq 1/3$. Further issues in this approach can be found in \cite{Davoudiasl:2004gf,Lambiase:2006md,Lambiase:2011by,Li:2004hh}.

 Another approach involves an axial vector current \cite{Debnath:2005wk,Mukhopadhyay:2005gb,Mukhopadhyay:2007vca,Sinha:2007uh}
instead of $J_{\mu}$. The scenario is based on the well known fact that fermions in curved
space-times exhibit a coupling of their spin to the curvature of the
background space-time.The Dirac Lagrangian density of a fermion can
be re-written as:
\begin{equation}\label{Bvector}
\mathcal{L}=\sqrt{-g}\,\overline{\psi}\left(i\gamma^{a}\partial_{a}-m+\gamma^{a}\gamma^{5}B_{a}\right)\psi~,\quad B^{d}=\epsilon^{abcd}e_{b\lambda}\left(\partial_{a}e_{\,\, c}^{\lambda}+\Gamma_{\nu\mu}^{\lambda}\, e_{\,\, c}^{\nu}\, e_{\,\, a}^{\mu}\right)~,
\end{equation}
in a standard notation, where $e_{\,\, a}^{\mu}$ are the vielbeins,
$\Gamma_{\,\alpha\beta}^{\mu}$ is the Christoffel connection and
Latin (Greek) letters denote tangent space (curved space-time) indices.
The space-time curvature background has, therefore, the effect of
inducing an ``axial'' background field $B_{a}$ which can be non-trivial
in certain anisotropic space-time geometries, such as Bianchi-type
cosmologies \cite{Debnath:2005wk,Mukhopadhyay:2005gb,Mukhopadhyay:2007vca,Sinha:2007uh}.
For an application to particle-antiparticle asymmetry it is necessary
for this axial field $B_{a}$ to be a constant in some local frame.
The existence of such a frame has not been demonstrated. As before
if it can be arranged that $B_{a}\neq0$ for $a=0$ then for constant
$B_{0}$ CPT is broken: the dispersion relation of neutrinos in such
backgrounds differs from that of antineutrinos. Explicitly we have
\begin{equation}
E=\sqrt{(\vec{p}-\vec{B})^{2}+m^{2}}+B_{0}~,\quad{\overline{E}}=\sqrt{(\vec{p}+\vec{B})^{2}+m^{2}}-B_{0}~.\label{nunubardr}
\end{equation}
 The relevant neutrino asymmetry emerges on following the same steps
(\emph{cf}. (\ref{cptvf}), (\ref{nnbarcptv})) used when there was
an explicit particle-antiparticle mass difference, As a consequence
the following neutrino-antineutrino density difference is found in
Bianchi II Cosmologies~\cite{Debnath:2005wk,Mukhopadhyay:2005gb,Mukhopadhyay:2007vca,Sinha:2007uh}:
\begin{equation}
\Delta n_{\nu}\equiv n_{\nu}-n_{\overline{\nu}}\sim g^{\star}\, T^{3}\left(\frac{B_{0}}{T}\right)\label{bianchi}
\end{equation}
 with $g^{\star}$ the number of degrees of freedom for the (relativistic) 
neutrino. An excess of particles over antiparticles is predicted only
when $B_{0}>0$, which had to be assumed in the analysis of \cite{Debnath:2005wk,Mukhopadhyay:2005gb,Mukhopadhyay:2007vca,Sinha:2007uh};
we should note, however, that the sign of $B_{0}$ and its constancy have not been justified in
this phenomenological approach~\footnote{The above considerations concern the dispersion relations for any fermion, not only neutrinos.
However, when one considers matter excitations from the vacuum, as relevant for leptogenesis, we need chiral fermions to get non trivial CPTV asymmetries in \emph{populations}
of particle and antiparticles, because $<\psi^\dagger \gamma^5 \psi> = - <\psi_L^\dagger \gamma^5 \psi_L> +  <\psi_R^\dagger \gamma^5 \psi_R> $.}.

At temperatures $T<T_{d}$, with $T_{d}$ the decoupling temperature
of the lepton-number violating processes, the ratio of the net Lepton
number $\Delta L$ (neutrino asymmetry) to entropy density (which
scales as $T^{3}$) remains constant,
\begin{equation}
\Delta L(T<T_{d})=\frac{\Delta n_{\nu}}{s}\sim\frac{B_{0}}{T_{d}}\label{dlbianchi}
\end{equation}
 which, for $T_{d}\sim10^{15}$~GeV and $B_{0}\sim10^{5}$~GeV,
implies a lepton asymmetry (leptogenesis) of order $\Delta L\sim10^{-10}$,
in agreement with observations. The latter can then be communicated
to the baryon sector to produce the observed BAU (baryogenesis), either
by a B-L conserving symmetry in the context of Grand Unified Theories
(GUT), or via (B\,+\,L)-conserving sphaleron processes, as in the
decoherence-induced CPT Violating case of \cite{Barenboim:2004ev,Barenboim:2004wu},
mentioned previously.

In closing this section, let us recapitulate what we regard as some of the less than satisfactory issues with the above proposals. The simple assumption of mass differences between particles and antiparticles dominated by the quark/antiquark mass differences cannot reproduce the observed BAU, because the latter are naturally bound by the current proton/anti-proton mass difference. Moreover, phenomenological models involving neutrino/antineutrino mass differences, although capable of reproducing the observed BAU, nevertheless are \emph{ad hoc} and lack microscopic justification. In addition, the models involving spin/curvature coupling require 
constancy of the axial vector $B^\mu$, which is assumed without proof. The same holds for the sign of the 
asymmetry between matter and antimatter in these models.  The models we will propose below address these problems in a more microscopic way and offer 
partial resolution. 

\section{CPTV-induced  in (String-Inspired) Background Geometry with Torsion}\label{sec:string}

In this section we would like to discuss an explicit case where a constant $B^0$ ``axial'' field arises as a consequence of the interaction of the fermion spin with a background geometry with \emph{torsion}. This is a novel result that, to the best of our knowledge, has not been noticed before. In the case of torsion the Christoffel symbol contains antisymmetric parts in its lower indices, so that $\Gamma^\lambda_{\,\,\,\,\mu\nu} \ne \Gamma^\lambda_{\,\,\,\,\nu\mu} $. As a consequence,  the last term of the right-hand side of the definition of the $B^d$ ``vector'' (\ref{Bvector}), which would vanish in a torsion free Einstein manifold, is \emph{not} zero. Such pure torsion terms may then contribute to CPTV dispersion relations, as we shall now demonstrate, even if the torsion-free metric does not.

The case we shall examine below is inspired from string theory.  It is known that  the massless gravitational multiplet of strings contains the dilaton (spin 0, scalar), $\Phi$, the
graviton (spin 2, symmetric tensor), $g_{\mu\nu} = g_{\nu\mu}$  and the spin 1 antisymmetric tensor $B_{\mu\nu} = - B_{\nu\mu}$ fields. The (Kalb-Ramond) field $B$ appears in the string effective action only through its totally antisymmetric field strength, $H_{\mu\nu\rho} = \partial_{\left[ \mu \right.} B_{\left.\nu\rho\right]}$, where  $[ \dots ]$ denotes antisymmetrization of the respective indices. It is known from string amplitude calculations~\cite{sloan} that $H_{\mu\nu\rho}$ plays the r\^ole of (Kalb-Ramond) \emph{torsion} in a generalised connection.
Indeed, the four-dimensional bosonic part of the $O(\alpha^\prime) $ effective action of the string (in the so-called Einstein frame, where the scalar curvature term is canonically normalised)
reads:
\begin{equation}\label{ea}
S= \frac{M_s^2 \, V^{c}}{16\pi} \int d^4x \sqrt{-g} \Big( R(g) - 2 \partial^\mu \Phi \partial_\mu \Phi - \frac{1}{12} e^{-4\Phi} H_{\mu\nu\rho} \, H^{\mu\nu\rho} + \dots \Big)
\end{equation}
where $M_s=1/\sqrt{\alpha^\prime}$ is the string mass scale and $V^{(c)}$ denotes the (dimensionless) compactification volume, where the compact radii are expressed in units of $\sqrt{\alpha'}$. From this form it becomes immediately obvious that the Kalb-Ramond tensor $H^2$ terms can be assembled together with the Einstein scalar curvature term $R(g)$
in a generalised curvature $\overline{R}(g, \overline{\Gamma})$ term defined with respect to
a generalised Christoffel symbol (connection) $\overline{\Gamma}$ with mixed symmetry in its lower indices
\begin{equation}\label{generalised}
\overline{\Gamma}^\lambda_{\,\,\,\mu\nu} = \Gamma^\lambda_{\,\,\,\mu\nu} + e^{-2\Phi} H^\lambda_{\mu\nu} \equiv
\Gamma^\lambda_{\,\,\,\mu\nu} + T^\lambda_{\,\,\,\mu\nu}~,
\end{equation}
where $\Gamma^\lambda_{\,\,\,\,\mu\nu} = \Gamma^\lambda_{\,\,\,\,\nu\mu} $ is the torsion-free Einstein-metric connection, and $T^\lambda_{\,\,\, \mu\nu} = - T^\lambda_{\,\,\,\nu\mu}$ is the \emph{torsion}.  It can be shown that this result persists in higher orders of $\alpha^\prime$, after appropriate field redefinitions, which leave the scattering amplitudes invariant.
Thus, for our purposes below, we consider it as an exact result, valid to all orders in stringy $\sigma$-model perturbation theory. Fermions then in such effective theories will couple to the $H$-tensor via spin connections with torsion, \emph{i.e.} the relevant Lagrangian terms (to lowest order in $\alpha^\prime$) will be of the form (\ref{Bvector})\footnote{We note that fermions coupled to Kalb-Ramond torsion tensors $H_{\mu\nu\rho}$
have been considered in the literature~\cite{kr1,kr2} but from a different perspective than ours.}.

In ref. \cite{aben} exact solutions to the conformal invariance conditions (to all orders in $\alpha^\prime$) of the low energy effective action of strings have been presented.  In four ``large'' (uncompactified)  dimensions of the string, the antisymmetric tensor field strength
can be written uniquely as
\begin{equation}\label{Hfield}
H_{\mu\nu\rho} = e^{2\Phi} \epsilon_{\mu\nu\rho\sigma} \partial^\sigma b (x)
\end{equation}
with $\epsilon_{0123} = \sqrt{g}$ and $\epsilon^{\mu\nu\rho\sigma} = |g|^{-1} \epsilon_{\mu\nu\rho\sigma}$, with $g$ the metric determinant.  The field
$b(x)$ is a ``pseudoscalar '' \emph{axion}-like field. It is worthy of mentioning that both the dilation $\Phi$ and axion $b$ fields are fields that appear as Goldstone bosons of spontaneously broken scale symmetries of the string vacua, and as such are exactly massless classically. In the effective string action such fields appear only through their derivatives hence a solution that is linear in time  for these two fields will only shift the various minima of all other fields in the effective action that couple to them by a space-time independent amount.

The exact solution of \cite{aben} is precisely such that in the string frame both dilation and axion fields are linear in target time $X^0$, $\Phi (X^0) \sim X^0$, $b(X^0) \sim X^0$.
In the ``physical''  Einstein frame (where cosmological observations are made) the temporal components of the
metric are normalised to $g_{00} =+1$
by an appropriate change of the time coordinate. In this setting,
the solution of \cite{aben} leads to  a linearly expanding Friedmann-Robertson-Walker (FRW) metric, with scale factor $a(t) \sim t$, with $t$ the FRW cosmic time. Moreover,  the dilaton field  $\Phi$ behaves as $- {\rm ln} t + \phi_0 $, with $\phi_0$ a constant, and the axion field $b(x)$ is  linear in time $t$, that is:
\begin{equation}\label{axion}
b(x) = \sqrt{2} e^{-\phi_0} \, \sqrt{Q^2} \,  \frac{M_s}{\sqrt{n}} t~,
\end{equation}
where $M_s$ is the string mass scale, $n$ is a positive integer, associated with the level of the Kac-Moody algebra of the underlying world-sheet  conformal field theory model and $Q^2 > 0 $  is its central-charge deficit (\emph{supercritical} string theory), with the central charge being given by: $c = 4 - 12 Q^2 - \frac{6}{n + 2} + c_I$, where $c_I$ is the central charge associated with the world-sheet conformal field theory of the compact ``internal'' dimensions of the string model~\cite{aben}. The requirement of cancellation of the world-sheet ghosts that appear as a result of fixing reparametrisation invariance of the world-sheet coordinates forces the constraint $c=26$. In the presence of a non-zero $Q^2 $ there is an additional  dark energy term in (\ref{ea}) of the form
$\int d^4 x \sqrt{-g} e^{2\Phi} (- 4Q^2)/\alpha^\prime $.
It should be noted that the linear axion field (\ref{axion}) remains a non-trivial solution of the string-effective-action equations \emph{even} in the \emph{static} space-time limit with a constant dilaton field~\cite{aben}.  In such a case the space time is an Einstein universe with positive cosmological constant and constant positive curvature proportional to
$6/(n+2)$.

In the above solutions, the covariant torsion tensor  $e^{-2\Phi} H_{\mu\nu\rho} $ is \emph{constant}, as becomes evident from (\ref{generalised}), (\ref{Hfield})
since the exponential  dilation factors cancel out in the relevant expressions. Only the spatial components of the torsion are non zero in this case,
\begin{equation}
T_{ijk} \sim \epsilon_{ijk} {\dot b} = \epsilon_{ijk} \sqrt{2 Q^2} e^{-\phi_0} \,  \frac{M_s}{\sqrt{n}}~,
\label{Hijk}
\end{equation}
where the overdot denotes time $t$ derivative.  From (\ref{generalised}), (\ref{Hfield}) and (\ref{Bvector}), we observe in this case that only the temporal component of the $B^d$ vector has non trivial, $B^0 \ne 0$,  and it is the $H$-torsion parts that contribute to it. The torsion-free gravitational parts on the other hand (for the FRW or flat case) yield vanishing contributions.
From (\ref{Bvector}) and (\ref{Hijk}) then we obtain a constant $B^0$ of order
\begin{equation}\label{b0string}
B^0 \sim \sqrt{2 Q^2} e^{-\phi_0} \, \frac{M_s}{\sqrt{n}} ~ {\rm GeV} > 0.
\end{equation}
Notice here that the sign of $B^0$ is fixed by string theory conventions.
From the previous discussion, to  get the phenomenologically acceptable leptogenesis in such a toy model it seems to require $B^0 \sim 10^{5} $~GeV. Since in string theory $M_s $ is a free parameter, restricted by phenomenological considerations to be higher than O(10$^4$) GeV, we thus see that we do not get a serious constraint on the Kac-Moody level from the requirement of CPTV-induced leptogenesis. Since the string coupling $g_s = e^{-\phi_0} < 1$ is assumed weak for phenomenological purposes ($g_s^2/4\pi \sim 1/20$), then we observe from (\ref{b0string}) that it is mainly the central charge deficit $Q^2$ of the underlying conformal field theory that determines the order of lepton asymmetries in this model.

\section{CPTV in Stochastic Finsler Geometries \label{sec:Finsler} }

Although all the models displaying CPTV that we have considered so far are based on local effective field theories, there is no compelling reason for a restriction to such a framework. In fact a microscopic model involving space-time defects based on string-brane theory suggests the use of a non-Riemannian metric background similar to that which occurs in Finsler geometry \cite{bao2000introduction,Girelli:2006fw}.(This model will be discussed in a subsequent section.) Independently there has been much interest in Finsler geometry~\cite{Vacaru:2002kp,Vacaru:2003uj,Bogoslovsky:1998wa,Bogoslovsky:2007gt,Bogoslovsky:2007mc,Gibbons:2007iu,Sindoni:2007rh,Anastasiei:2007dx,Vacaru:2012ve,Vacaru:2010fa} for characterising the Early universe \cite{Kouretsis:2008ha,Magueijo:2002xx,Kouretsis:2010vs,Stavrinos:2012ty,Vacaru:2010fi} and for descriptions of modified dispersion relations for particle probes~\cite{Ellis:1999uh,Magueijo:2002xx,Girelli:2006fw,Skakala:2008kf}.
Finsler  geometry  has a metric which, in addition to space-time coordinates, depends also on ``\emph{velocities}''. Lorentz symmetry is broken through some fixed vectors in the metric. We explore the consequences of making such vectors having components which are stochastic with possibly zero mean. This is a feature that arises in the defect model of D-foam~\cite{Ellis:1999uh, Ellis:2004ay, Ellis:2005ib} that we will consider in the next section. The defects stochastically fluctuate, due to both statistical and quantum stringy effects in large populations of such D-particles that can populate eras of the early universe. As we shall discuss below, the result of the interaction of neutrinos  with these defects, leads to stochastically fluctuating Finsler-like metrics. However we wish to consider the consequences for CPTV and matter-antimatter asymmetry of this stochasticity in a general context.
This underlying model provides our main motivation to study this class of space-times in this section and to contrast our findings on the induced CPTV for such cases with the corresponding ones for the D-foam model. For clarity, we commence our discussion with a brief reminder of the definition and properties of Finsler geometries, and then we proceed to discuss CPTV issues in a particular, but representative, Finsler-like geometry, which however is stochastically fluctuating.

A Finsler geometry on a manifold $M$ is defined in terms of a Finsler
norm $F\left(x,y\right)$, a real function of two arguments $x$ and $y$, where $x\in M$ and $y\in T_{x}M$
(the tangent space at $x$) \cite{bao2000introduction,Girelli:2006fw}. The norm $F\left(x,y\right)$ satisfies:
\begin{itemize}
\item $F\left(x,y\right)\neq0$ if $y\neq0$
\item $F\left(x,\lambda y\right)=\left|\lambda\right|F\left(x,y\right)$for
$\lambda\in\mathbb{R}$.
\end{itemize}
The Finsler geometry is defined in terms of a metric $g_{\mu\nu}\left(x,y\right)$ which
is given in terms of the Finsler norm
\begin{equation}
g_{\mu\nu}\left(x,y\right)\equiv\frac{1}{2}\frac{\partial^{2}F^{2}\left(x,y\right)}{\partial y^{\mu}\partial y^{\nu}}.\label{Finsler metric}
\end{equation}
The inverse of $g_{\mu\nu}\left(x,y\right)$ is represented by $g^{\mu\nu}\left(x,y\right)$.
It is necessary to define a similar structure in phase space (i.e.
for the co-tangent space). The dual $\omega_{\mu}\left(y\right)$
to $y^{\mu}$is defined by \cite{Girelli:2006fw}
\begin{equation}
\omega_{\mu}\left(y\right)=g_{\mu\nu}\left(x,y\right)y^{\nu}.\label{dual}
\end{equation}
Furthermore this relation can be inverted so that corresponding to
$\omega$ there is a dual $y\left(\omega\right).$ The Finsler norm
$G\left(x,\omega\right)$on cotangent space is defined by
\begin{equation}
G\left(x,\omega\right)=F\left(x,y\left(\omega\right)\right).\label{form norm}
\end{equation}
The Finsler metric $h^{\mu\nu}\left(x,\omega\right)$ in phase space
is defined analogously to the usual Finsler metric in (\ref{Finsler metric})
\begin{equation}
h^{\mu\nu}\left(x,\omega\right)=\frac{1}{2}\frac{\partial^{2}G^{2}\left(x,\omega\right)}{\partial\omega_{\mu}\partial\omega_{\nu}}.\label{dual metric}
\end{equation}
It is interesting to note that Cartan's torsion tensor $C_{\mu\nu\delta}=\frac{1}{2}\frac{\partial g_{\mu\nu}}{\partial y^{\delta}}$
vanishes when $g_{\mu\nu}$ is Riemannian. Much of the mathematical
literature has dealt with Finsler extensions of Riemannian geometry
when the metric signature has been euclidean. However we need to consider
pseudo-Riemannian structures. This can be formally done. However the
Finsler norm leads to a metric which can lead to singularities in
the metric for off-shell test particles.

We focus on on-shell neutrinos
(which are now known to have small masses). Again, this is motivated by our desire to discuss leptogenesis in such geometries.
Moreover, for reasons that will become clear in section \ref{sec:dfoam}, it is neutrinos that play a preferential r\^ole in interacting non-trivially with the D-particle foam background,
which induces stochastically fluctuating Finsler-like space times. As our main motivation is to compare the generic Finsler-like case with the D-foam model, as far as CPTV is concerned, we restrict our attention here on the effects of stochastically fluctuating Finsler geometries on dispersion relations of neutrinos and antineutrinos.
We shall consider a particular
type of Finsler metric on a manifold $M$ which is known as the Randers
metric \cite{bao2000introduction}~\footnote{We mention, for completeness, that
the other popular class of Finsler geometries, that appears in the General Relativistic version~\cite{Bogoslovsky:1998wa,Gibbons:2007iu,Bogoslovsky:2007gt,Bogoslovsky:2007mc} of the so-called Very Special Relativety Model~\cite{VSR}, and cosmological extensions thereof~\cite{Kouretsis:2008ha,Kouretsis:2010vs} are \emph{not } characterised by CPTV in the dispersion relations, nor of the spin-curvature type discussed in section \ref{sec:bg}. In fact such VSR-related models have been prop[osed in the past as candidates for the generation of Lepton-number conserving neutrino masses~\cite{cgneu}, and hence our ~Lepton-number violating considerations in this work do not apply.}. The norm $F_{R}$ for the Randers
metric is
\begin{equation}
F_{R}\left(x,y\right)=\alpha\left(x,y\right)+\beta\left(x,y\right)\label{randers1}
\end{equation}
where
\begin{equation}
\alpha\left(x,y\right)=\sqrt{r_{\mu\nu}\left(x\right)y^{\mu}y^{\nu}}\label{randers2}
\end{equation}
and
\begin{equation}
\beta\left(x,y\right)=b_{\mu}\left(x\right)y^{\mu}.\label{randers3}
\end{equation}
The conditions for a Finsler norm are satisfied. It was noted in \cite{Gibbons:2008zi}
that the geodesics of this metric coincided with the minimum time
trajectories of a particle moving on a Riemannian manifold in the
presence of a time independent drift given by a vector field. This
is similar to Fermat's principle for propagation in refractive media. Similarities
of D-particle foam to a refracting medium~\cite{Ellis:1999uh,Ellis:2008gg,Li:2009tt}, will be mentioned briefly
in the next section. If we were to assume that the result on minimum
time trajectories was true for a pseudo-Riemannian situation and the
drift was given by collisions due to D-particle scattering, then at
a heuristic level a stochastic drift could be a reasonable generic
phenomenological model of the back-reaction of low dimensional recoiling
branes on matter. We shall write
\begin{equation}
b_{\mu}\left(x\right)=\phi\left(x\right)l_{\mu}\label{randers4}
\end{equation}
where $l_{\mu}$is a constant vector. In our model $\phi\left(x\right)$ will
be a gaussian stochastic variable. On average the metric will be like
a Riemannian metric if the mean of $\left\langle \phi\right\rangle $ vanishes.
From (\ref{Finsler metric}) we deduce that
\[
\begin{split}
g_{\mu\nu}\left(x,y\right)&  =r_{\mu\nu}\left(x\right)+\phi^{2}\left(x\right)l_{\mu}l_{\nu}+\left(\frac{r_{\mu\nu}\left(x\right)}{\alpha\left(x,y\right)}-\frac{r_{\mu\varrho}\left(x\right)r_{\nu\sigma}\left(x\right)y^{\varrho}y^{\sigma}}{\alpha\left(x,y\right)^{3}}\right)\phi\left(x\right)l_{c}y^{c}\\
& \quad +\frac{1}{\alpha\left(x,y\right)}\left(r_{\mu\varrho}\left(x\right)y^{\varrho}\phi\left(x\right)l_{\nu}+r_{\nu\varrho}\left(x\right)y^{\varrho}\phi\left(x\right)l_{\mu}\right).\\
\end{split}
\]
We shall consider now a situation with $r_{\mu\nu}\left(x\right)=\eta_{\mu\nu}$ where
$\eta_{\mu\nu}$ is the diagonal Minkowski matrix with entries $\left(1,-1,-1,-1\right)$.
(The summation convention of repeated indices will be always understood
unless explicitly stated otherwise.) Within the framework of a Robertson-Walker
metric we shall ignore effects on the time-scale of the inverse expansion
rate. Let us introduce some notation:
\[
\widetilde{\alpha}=\sqrt{\eta_{\mu\varrho}y^{\mu}y^{\varrho}},
\]

\[
\widetilde{\beta}=l_{\mu}y^{\mu},
\]
and
\[
C_{\mu}=y^{\nu}\eta_{\nu\mu}.
\]
We can then rewrite $g_{\mu\nu}\left(x,y\right)$as
\[
g_{\mu\nu}=\eta_{\mu\nu}+\phi^{2}l_{\mu}l_{\nu}+\left(\frac{\eta_{\mu\nu}}{\widetilde{\alpha}}-\frac{C_{\mu}C_{\nu}}{\widetilde{\alpha}^{3}}\right)\phi\widetilde{\beta}+\left(C_{\mu}l_{\nu}+C_{\nu}l_{\mu}\right)\phi\widetilde{\alpha}^{-1}.
\]
However in order to construct energy-momentum dispersion relations
we will need to calculate the dual metric $h^{\mu\nu}$ (\emph{cf}. (\ref{dual metric}))
which involves $y^{\mu}\left(\omega\right).$ We shall consider $\phi$ to
be small and work to lowest order in the evaluation of $G\left(x,\omega\right)$.
First we write $y^{\mu}=y_{\left(0\right)}^{\mu}+\phi y_{\left(1\right)}^{\mu}$.
From (\ref{dual}) we have
\[
\omega_{\mu}=g_{\mu\nu}\left(x,y_{\left(0\right)}+\phi y_{\left(1\right)}\right)\left(y_{\left(0\right)}^{\nu}+\phi y_{\left(1\right)}^{\nu}\right).
\]
 We can solve for $y_{\left(0\right)}$and $y_{\left(1\right)}.$We
find for $y_{\left(0\right)}$
\[
y_{\left(0\right)}^{0}=-\omega_{0},\, y_{\left(0\right)}^{1}=\omega_{1},\, y_{\left(0\right)}^{2}=\omega_{2},\, y_{\left(0\right)}^{3}=\omega_{3}.
\]
We find for $y_{\left(1\right)}^{\mu}$
\[
y_{\left(1\right)}^{\mu}=\frac{l_{\mu}\varpi^{2}-\omega_{\mu}d}{\varpi}
\]
where $\varpi=\sqrt{\eta^{\kappa\delta}\omega_{\kappa}\omega_{\delta}}$
and $d=l_{\kappa}\omega_{\kappa}$. Hence
\begin{equation}
G\left(x,\omega\right)=\varpi-\frac{\phi}{\varpi^{2}}l_{\kappa}\omega_{\kappa}\left(\varpi^{2}+2\omega_{0}^{2}\right).\label{form norm2}
\end{equation}
We deduce that
\begin{equation}
h^{\mu\nu}=\eta^{\mu\nu}+\phi H^{\mu\nu}\label{dual metric2}
\end{equation}
where

\[
H^{00}=\frac{-l_{\kappa}\omega_{\kappa}(5\varpi^{4}+11\varpi^{2}\omega_{0}^{2}+6\omega_{0}^{4})-2l_{0}\omega_{0}\left(5\varpi^{4}+2\varpi^{2}\omega_{0}^{2}\right)}{\varpi^{5}},
\]
for $j=1,2,3$
\[
H^{jj}=\frac{1}{\varpi^{5}}\left[2l_{j}\omega_{j}\varpi^{2}\left(\varpi^{2}+2\omega_{0}^{2}\right)+l_{\kappa}\omega_{\kappa}\left(3\varpi^{2}\left(\varpi^{2}-\omega_{j}^{2}\right)-2\omega_{0}^{2}\left(\varpi^{2}-3\omega_{j}^{2}\right)\right)\right],\;
\]
\[
H^{0j}=\frac{1}{\varpi^{5}}\left[-l_{j}\left(5\varpi^{4}\omega_{0}+2\varpi^{2}\omega_{0}^{3}\right)+\omega_{j}\left(l_{0}\left(\varpi^{4}+2\varpi^{2}\omega_{0}^{2}\right)+l_{\kappa}\omega_{\kappa}\omega_{0}\left(5\varpi^{2}+6\omega_{0}^{2}\right)\right)\right]
\]
and for $\: i<j\; for\: i=1,2,3$
\[
\begin{split}
H^{ij}& =\frac{\omega_{i}l_{j}\left(\varpi^{2}+2\omega_{0}^{2}\right)}{\varpi^{3}}\\
& \quad +\omega_{j}\left\{ \frac{l_{i}\left(-\varpi^{4}-6\omega_{0}^{2}\omega_{i}^{2}+\varpi^{2}\left(2\omega_{0}^{2}+\omega_{i}^{2}\right)\right)}{\varpi^{5}}+\frac{\left(\varpi^{2}-6\omega_{0}^{2}\right)\omega_{i}\left(l_{\delta}\omega_{\delta}-l_{i}\omega_{i}\right)}{\varpi^{5}}+\frac{2\omega_{j}}{\varpi^{4}}\left(l_{i}\varpi^{2}-\omega_{i}l_{\delta}\omega_{\delta}\right)\right\} .
\end{split}
\]

We have assumed a homogeneous $\phi$ with $\phi$ being $x$ independent.
The mass shell condition $h^{\mu\nu}\omega_{\mu}\omega_{\nu}=m^{2}$ leads
to the following equation for $\omega_{0}$:
\begin{equation}
\varpi^{2}-\frac{2\phi}{m}\left\{ l_{0}\left[\omega_{0}^{3}+\omega_{0}\vec{\omega}^{2}\right]+l_{1}\left[\omega_{1}\left(\omega_{0}^{2}+\vec{\omega}^{2}\right)\right]+l_{2}\left[\omega_{2}\left(\omega_{0}^{2}+\vec{\omega}^{2}\right)\right]+l_{3}\left[\omega_{3}\left(\omega_{0}^{2}+\vec{\omega}^{2}\right)\right]\right\} =m^{2}\label{mass_shell}
\end{equation}
where $\vec{\omega}^{2}\equiv\omega_{1}^{2}+\omega_{2}^{2}+\omega_{3}^{2}.$

In the model it is possible to choose $l_{\mu}.$ Not all choices will
lead to asymmetric population distributions between particles and
anti-particles.The space-like choice $l_{0}=0$ gives a degenerate
spectrum for particle and anti-particle and hence no CPTV in dispersion relations:
\begin{equation}
\omega_{0}=\left[\vec{\omega}^{2}\left(1+\frac{2\phi}{m}\left(l_{1}\omega_{1}+l_{2}\omega_{2}+l_{3}\omega_{3}\right)\right)+m^{2}\right]^{1/2}\left[1-\frac{2\phi}{m}\left(l_{1}\omega_{1}+l_{2}\omega_{2}+l_{3}\omega_{3}\right)\right]^{-1/2}.\label{spacelike}
\end{equation}
Therefore this case cannot be used for Leptogenesis in our framework.

More generally the dispersion relation is
\begin{equation}
\omega_{0}=\pm\frac{\phi}{m}l_{0}\left(2\vec{\omega}^{2}+m^{2}\right)+\mathfrak{K}\left(\phi,\omega,m\right)\label{FinslerDisp1}
\end{equation}
where
\[
\mathfrak{K}\left(\phi,\omega,m\right)=\left(\vec{\omega}^{2}+m^{2}+\frac{2\phi}{m}\left(2\vec{\omega}^{2}+m^{2}\right)\left(l_{1}\omega_{1}+l_{2}\omega_{2}+l_{3}\omega_{3}\right)\right)^{1/2}.
\]
The $+$sign is for the particle and the $-$sign is for the antiparticle.
For the ``time-like'' case $l_{0}=1,\, l_{1}=l_{2}=l_{3}=0$ the
dispersion relation reduces to e
\begin{equation}
\omega_{0}=\sqrt{\vec{\omega}^{2}+m^{2}}\pm\frac{\phi}{m}\left(2\vec{\omega}^{2}+m^{2}\right).\label{FinslerDisp2}
\end{equation}
The sign of $l_{0}$ can be reabsorbed in $\phi$.

For the ``null'' case $l_{0}=l_{1}=1$ and $l_{2}=l_{3}=0$ the dispersion
relation reduces to
\begin{equation}
\omega_{0}=\sqrt{\vec{\omega}^{2}+m^{2}}\left(1+\frac{\phi}{m}\left(2-\frac{m^{2}}{\vec{\omega}^{2}+m^{2}}\right)\omega_{1}\right)\pm\frac{\phi}{m}\left(2\vec{\omega}^{2}+m^{2}\right).\label{FinslerDisp3}
\end{equation}
 The parameters ${\vec \omega}$ play the r\^ole of momenta $\vec{p}$ in our case of neutrinos of mass $m = m_\nu$ propagating in these space-times, and the Finsler metric may be seen as sort of back reaction on the space-time of such a propagation (to better appreciate this, the reader is invited to the discussion in the next section \ref{sec:dfoam}, where a particular model of D(efect)-foam is considered as a medium for neutrino propagation in the early universe, leading to Finsler-like metric distortions as a consequence of medum/particle interactions).

Corresponding to such
models involving D-foam, the parameter $\phi$ is modelled as a stochastic gaussian
process with a mean $\mathfrak{a}$ and standard deviation $\sigma$.
The interesting features in earlier analyses~\cite{Bernabeu:2006av},\cite{Mavromatos:2010nk}
were retained when $\mathfrak{a}=0$. The fermion number distribution
$n$ from equilibrium statistical mechanics is given by
\begin{equation}
n= \, g_{d.o.f.} \, \int\frac{d^{3}p}{\left(2\pi\right)^{3}}\frac{1}{\exp\left(\beta\left(\omega_{0}-\mu\right)\right)+1}\label{numberDistribn}
\end{equation}
where we have ignored degeneracy factors. In terms of spherical polars
$d^{3}p=d\xi {\rm sin}\theta \, d\theta dp\, p^{2}$ where $p=\left|\vec{p}\right|,$
$\theta$ lies in $\left[0,\pi\right]$ and $\xi$ lies in $\left[0,2\pi\right]$.
 In the regime of validity of our analysis (\emph{i.e}. relativistic neutrinos ($\mu \simeq 0$) that decouple at high temperatures $T \gg m_\nu $ and for small stochastic fluctuations of the background geometry)
we will consider $\beta = 1/T$ (in units of the Boltzmann constant $k_B =1$) small and expand the relevant expressions in a power series. We denote $n$ averaged over the
distribution of $\phi$ as $\ll n\gg$ . It is given by
\begin{equation}
\ll n\gg\equiv  \, g_{d.o.f.} \, \int_{-\infty}^{\infty}d\phi\frac{\exp\left(-\frac{\left(\phi-\mathfrak{a}\right)^{2}}{\sigma^{2}}\right)}{\sqrt{\pi}\sigma}n.\label{Avgnumber}
\end{equation}

First of all, it immediately follows from (\ref{FinslerDisp2}), that for the ``time-like'' case, when $\mathfrak{a} = 0$, there is no particle/antiparticle asymmetry.
This is to be expected, given that $\mathfrak{a} \ne 0$ corresponds in a sense to an averaged  Lorentz violation in this
stochastic geometry, and hence one of the basic assumptions for CPT Invariance of the effective theory of neutrinos in this ``medium'' is relaxed.

For the ``time-like'' case, when $\mathfrak{a} \ne 0$ and  $\beta$ small,
we obtain to leading order in $T/m \gg 1$:
\begin{equation}
\ll\triangle n\gg \sim  -  \frac{2}{\pi^2} \mathfrak{a} \, g_{d.o.f.} T^3 \,  \Big( \frac{T}{m}\Big) \int_0^\infty \frac{dx \, x^4 \, e^x}{(1 + e^x)^2} =  -  \mathfrak{a} \, g_{d.o.f.} T^3 \frac{7 \pi^2}{15}  \, \Big( \frac{T}{m}\Big)
\label{timelike}
\end{equation}
We require $\mathfrak{a}<0$ in order to have a particle-antiparticle
asymmetry where the particle distribution dominates the antiparticle
distribution. This yields the following Lepton (neutrino)  asymmetry, assumed to freeze at the neutrino decoupling temperature  $T_d \sim 10^{15}$ GeV
\begin{equation}\label{finslerdl}
\triangle L (T \sim T_d) = \frac{\triangle n_\nu }{s} \sim - 10  \, \mathfrak{a} \,  \, \frac{T_d}{m_\nu}
\end{equation}
where, as usual, $s$ denotes the entropy density, which for relativistic species is assumed to be $s \sim g_{d.o.f.} \frac{2\pi^2}{45} \, T^3 $.

To obtain the phenomenologically correct value of $\Delta L (T \sim T_d) \sim 10^{-10}$, which is then communicated to the baryon sector via B+L violating sphaleron processes, or B-L conserving grand unified models (assumed appropriately embedded in such space-time geometries), one needs to take into account that, according to current data, the masses of the active neutrinos that  are assumed to participate in (\ref{finslerdl}) must be smaller than $m_\nu < 0.2$~eV. This implies then that one needs only an extremely small in magnitude violation of Lorentz symmetry on average in this stochastic Finsler space time,  $\mathfrak{a} \sim - 10^{-36} $, in order to reproduce the observed Baryon Asymmetry in the universe.
The assumption of fixing the sign of $\mathfrak{a}$ is considered as fine tuning, and is a feature that is common in  the models of gravitational leptogenesis/baryogenesis that exist  in the current literature, as discussed briefly above~\cite{Dvali:1999gf,Davoudiasl:2004gf,Debnath:2005wk,Mukhopadhyay:2005gb,Mukhopadhyay:2007vca,Sinha:2007uh,Lambiase:2006md,Lambiase:2011by,Li:2004hh}.

Let us next calculate the asymmetry for the null case (\ref{FinslerDisp3}).
For $\mathfrak{a}=0$ there is no asymmetry as we can see from considering
(\ref{Avgnumber}) where we will exchange the order of the integrations
over $\vec{p}$ and $\phi$ . The $\phi$ averaged expression  for the \textit{particle} distribution leads to

\begin{equation}\label{nullparticle}
  \frac{1}{\sigma\sqrt{\pi}}\int_{-\infty}^{\infty}d\phi\,\exp\left(-\frac{(\phi-\mathfrak{a})^{2}}{\sigma^{2}}\right)\frac{T^{3}}{(2\pi)^{2}}\int_{0}^{\infty} dx \, x^{2} \int_{0}^{\pi} d\theta \sin \theta  \frac{1}{1+ e^{x}}(1+ A(x,\theta))
\end{equation}

where
\[
\begin{split}
A(x,\theta)& =\{-\frac{e^{x}}{1+e^{x}}[\frac{2\phi}{\epsilon}x^{2}(1+\cos \theta)+\frac{2\phi^{2}}{\epsilon^{2}}x^{4}(1+\cos \theta)^{2}]\\
& \quad +\frac{e^{2x}}{(1+e^{x})^{2}}\frac{4\phi^{2}}{\epsilon^{2}}x^{4}(1+ \cos \theta)^{2}\}
\end{split}
\]
we have expressed $\vec{p}$ in terms of spherical polars, $x= \frac{|\overrightarrow{p}|}{T}$ and
the $\theta=0$ axis has been taken for convenience to coincide with
the 1-axis;  The corresponding expression for the \textit{antiparticle} is
\begin{equation}\label{nullantipart}
  \frac{1}{\sigma\sqrt{\pi}}\int_{-\infty}^{\infty}d\phi\,\exp\left(-\frac{(\phi-\mathfrak{a})^{2}}{\sigma^{2}}\right)\frac{T^{3}}{(2\pi)^{2}}\int_{0}^{\infty} dx \, x^{2} \int_{0}^{\pi} d\theta \sin \theta  \frac{1}{1+ e^{x}}(1+ B(x,\theta))
\end{equation}
where
\[
\begin{split}
B(x,\theta)& =\{-\frac{e^{x}}{1+e^{x}}[-\frac{2\phi}{\epsilon}x^{2}(1-\cos \theta)+\frac{2\phi^{2}}{\epsilon^{2}}x^{4}(1-\cos \theta)^{2}]\\
& \quad +\frac{e^{2x}}{(1+e^{x})^{2}}\frac{4\phi^{2}}{\epsilon^{2}}x^{4}(1- \cos \theta)^{2}\}
\end{split}
\]
Although $A(x,\theta)\neq B(x,\theta)$, for   $\mathfrak{a}=0$,   on
integrating over $\theta$ (as part of the integration over $\vec{p}$)
we have
\[
\ll\int_{0}^{\pi}d\theta\sin\theta\: A(x,\theta) \gg \,= \, \ll \int_{0}^{\pi}d\theta\sin\theta\: B(x,\theta) \gg.
\]
Hence when $\mathfrak{a}=0$ there is no particle-antiparticle asymmetry.
Hence we need Lorentz violation in the mean to obtain leptogenesis.
This contrasts with our earlier work on correlations in neutral meson
pairs created in meson factories where a signature of CPT violation~\cite{Bernabeu:2006av}
was present even for $\mathfrak{a}=0$.

When $\mathfrak{a} \neq 0$ the counterpart
of (\ref{nullparticle}) is
\begin{equation}
\frac{2}{\sigma\sqrt{\pi}}\int_{-\infty}^{\infty}d\phi\,\exp\left(-\frac{(\phi-\mathfrak{a})^{2}}{\sigma^{2}}\right)\left( \frac{1}{1+e^{x}}\{x^{2}-\frac{2\phi}{\epsilon}\frac{x^{4}e^{x}}{1+e^{x}}\} \right)\label{nullpart1}
\end{equation}
and the counterpart of (\ref{nullantipart}) is
\begin{equation}
\frac{2}{\sigma\sqrt{\pi}}\int_{-\infty}^{\infty}d\phi\,\exp\left(-\frac{(\phi-\mathfrak{a})^{2}}{\sigma^{2}}\right)\left( \frac{1}{1+e^{x}}\{x^{2}+\frac{2\phi}{\epsilon}\frac{x^{4}e^{x}}{1+e^{x}}\} \right).\label{nullantipart1}
\end{equation}

Hence the analysis reduces to that for the time-like case. Consequently
we have again (\ref{timelike}) and require $\mathfrak{a} < 0$ to obtain an asymmetry
with particle distribution exceeding that for the anti-particle.

This need to  tune the sign of the asymmetry in the approaches described so far leads us to pursue a different route to relations of the
form (\ref{nunubardr}) in the remainder of this  paper . The route is not based on an approach involving
effective local field theory. The framework we will follow
fits naturally into a picture which has been advocated in the past
to understand gravitational decoherence \cite{Mavromatos:2005bu,Bernabeu:2006av} and dark matter abundance \cite{Mavromatos:2010jt}.
and may be considered to be a more microscopic approach than the earlier proposals thatwe review.
For D-foam the asymmetry is controlled
by a parameter whose sign does not need adjustment. Furthermore, the
constancy of the parameter, which has a stochastic origin, seems to
be a more reasonable assumption than the same requirement for curvature.
The D-particle foam model also has an interesting feature in that
a non-Riemmanian metric is induced on the motion of matter on the
brane world. Our mechanism of CPTV in D-foam did not directly rely
on this aspect. However since this non-Riemannian structure is somewhat
similar to the structure of Finsler metrics (see e.g. \cite{Girelli:2006fw}
and references therein) we have explored the implications for CPTV
within the more general framework of stochastic Finsler metrics. We
will find that it is indeed possible to have CPTV but in general it
is necessary to choose the appropriate sign of s the mean of a stochastic
parameter.

\section{Stringy-Defect(D-)foam-induced CPTV and Leptogenesis \label{sec:dfoam} }

In this section we shall consider a population of D0-branes (or lower dimensional compactified D-branes which are effectively point-like from from the point of view of a brane world observer \footnote{It should be remarked that for the effective compactified D-``particles'' the interactions with the charged matter excitations are suppressed relative to the neutral ones~\cite{Li:2009tt}. Hence, even in this case,  it is the electrically neutral excitations which interact primarily with the D-foam.}) interacting with neutral fermions such as the neutrino and anti-neutrino. This interaction leads to different dispersion relations for neutrinos and anti-neutrinos which in turn leads to an excess of the population of neutrinos over anti-neutrinos. The freeze-out of neutrinos at the decoupling temperature of neutrinos leads to leptogenesis given by the standard cosmological considerations.
The latter results, through standard Baryon (B) and Lepton (L) number violating sphaleron processes or B-L conserving interactions in grand unified models describing matter excitations on the
brane, to the observed Baryon Asymmetry in the Unverse, with complete  dominance of matter over  antimatter, in a rather natural way, as we shall discuss below.
 Moreover, as we shall explain below, in this model of D(efect)-foam, the prevalence of matter over antimatter,\emph{ i.e.} the \emph{positive sign} of the asymmetry $\Delta n >0$, follows naturally, as a consequence of loss of energy of neutrinos during their interactions  with the space-time defects, due to recoil of the latter.  Thus, the sign of the induced asymmetry need not be fixed by hand, unlike the cases of gravitational leptogenesis discussed in previous sections.
For instructive purposes, we first discuss the properties of the foam model, in the next subsection, before moving onto issues of CPTV and leptogenesis.
\begin{figure}[ht]
\centering \includegraphics[width=5.5cm]{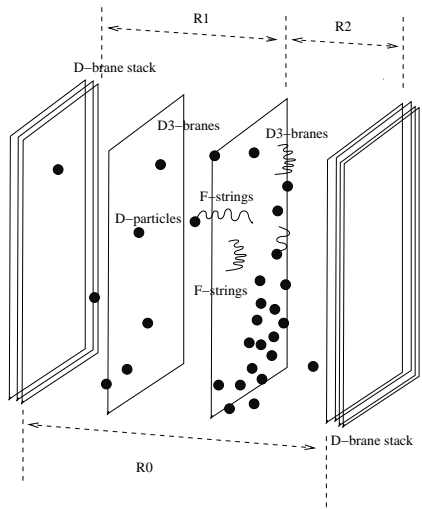} \caption{Schematic representation of a generic D-particle space-time foam model.
The model of ref.~\cite{Ellis:2004ay,Ellis:2005ib}, which acts as
a prototype of a D-foam, involves two stacks of D8-branes, each stack
being attached to an orientifold plane. Owing to their special reflective
properties, the latter provide a natural compactification of the bulk
dimension. The bulk is punctured by D0-branes (D-particles), which
are allowed in the type~IA string theory. The presence of a D-brane
is essential due to gauge flux conservation, since an isolated D-particle
cannot exist. Open (F-)strings live on the brane world, representing
Standard Model (SM) matter and they can interact in a topologically
non-trivial way with the D-particle defects in the foam, only if they
do not carry electric flux (electrically neutral excitations). Thus,
from the SM matter excitations on the brane, mainly neutrinos interact
non-trivially with the D-foam. Recoil of the D-particle during such
interactions creates appropriate distortion in the space-time geometry,
which depend on the momenta of the incident string states, and thus
are of a generalized Finsler type. The propagation of neutrinos in
such geometries is CPT Violating.}

\label{fig:recoil}
\end{figure}

\subsection{The D-foam Model of the universe and Neutrinos} \label{sec:dfoamprop}

D-foam models~\cite{Ellis:2005ib,Ellis:2004ay,Bernabeu:2006av} are
stringy models of space-time foamy geometries, which involve brane
universes, propagating in higher-dimensional bulk geometries. The
bulk contains point-like D-brane defects (``D-particles'' or D$0$ branes) whose
population density is constrained by the amount of CPTV that we observe. In many
string theories (such as bosonic and type~IIA string theories) they are stable
zero-dimensional defects. However for our purposes we will
consider them to be present in string theories of phenomenological
interest~\cite{Ellis:2008gg} since, even when elementary D-particles
cannot exist consistently, as is the case of type~IIB string models,
there can be effective D-particles formed by the compactification
of higher dimensional D-branes~\cite{Li:2009tt} (\emph{e.g}. three-branes
wrapped around three-cycles, with relatively small radii).
In general the construction
of a model involves a number of parallel brane worlds with three large
spatial dimensions, the required number being determined by target
space SUSY.(Phenomenologically realistic models may require stacks of
intersecting branes arranged in particular ways \cite{Zwiebach:789942}.)
These brane worlds move in a bulk space-time containing a gas of point-like
bulk branes, called D-particles, which are stringy space-time solitonic
defects~\cite{Duff:1994an} (\emph{cf.} fig.~\ref{fig:recoil}). One of these branes
is the observable universe. On this brane the D-particles will appear
as space-time defects. Typically open strings interact with D-particles
and satisfy Dirichlet boundary conditions when attached to them. Closed
and open strings may be ``cut'' by D-particles, a process that involves
capture of the incident open string and creation of stretched strings
between the (recoiling) D-particle and the brane world (string ``splitting''),
and subsequent re-emission of the open string.
It has also been speculated that nucleation of localised compactified defects \cite{Dvali:1999gf} from
such a D brane world (at the very high temperatures in the early universe)
can be considered as a generation of compactified effective (metastable \cite{Linde:1981zj} but very long lived) D-particles.

The preferential r\^ole of neutrinos in feeling the full effects of
D-foam, and hence the CPTV, as we shall discuss below, is attributed
to electric charge conservation: the representation of SM particles
as open strings, with their ends attached to the brane worlds, prevents
capture and splitting of open strings carrying electric fluxes by
the D-particles. (We should recall that in string theory the electric
charge is at the end point of an open string.) D-particles are electrically
neutral and thus electric charge would not have been conserved if
such processes had taken place. This is also consistent with the effective
D-particles which may have formed as a result of nucleation \cite{Dvali:1999gf}. Hence,
the D-particle foam is transparent to charged excitations of the SM,
leaving neutral particles, in particular neutrinos, susceptible
to the foam effects.

As discussed in detail in \cite{Mavromatos:2010nk} the density of
D-particles on the brane world is permitted to be relatively large, even at late
eras of the universe, given the fact that bulk D-particles exert
forces on the brane universe with mixed sign contributions to the
brane vacuum energy, depending on the distance of the bulk D-particles
from the brane \cite{Mavromatos:2009rf,Douglas:1996yp}. Such forces
are due to stretched strings between the
defect and the brane. These energy contributions depend only on the
transverse components of the relative velocities
of the defect with respect to the brane worlds.
In fact, depending on the distance of the bulk D-particle from the brane world,
the sign of the contributions on the D-brane vacuum energy from the moving defect in the bulk,
with velocity
$v$ perpendicular to the brane world,
may be negative or positive.
In particular, the interaction of a single D-particle,
that lies far away from the D3 brane (D8-compactified) world, and
moves adiabatically with a small velocity $v_{\perp}$ in a direction
transverse to the brane, results in the following potential~%
\footnote{For brevity, in what follows we ignore potential contributions induced
by compactification of the D8 brane worlds to D3 branes, stating only
the expressions for the induced potential on the uncompactified brane
world as a result of a stretched string between the latter and the
D-particle - the compactitication does not affect our arguments on
the negative energy contributions to the brane vacuum energy.%
}~\cite{Ellis:2005ib}
\begin{eqnarray}
\mathcal{V}_{D0-D8}^{long}=+\frac{r\,(v_{\perp}^{{\rm long}})^{2}}{8\pi\alpha^{\prime}}~,~r\gg\sqrt{\alpha^{\prime}}.\label{long-1}
\end{eqnarray}
On the other hand, a D-particle close to the D3-brane (compactified
D8), at a distance $r'\ll\sqrt{\alpha'}$, moving adiabatically in
the perpendicular direction with a velocity $v_{\perp}^{{\rm short}}$
will induce the following potential to it:
\begin{equation}\label{pot1} \mathcal{V}_{D0-D8}^{short}=
- \frac{\pi\alpha^\prime (v_\perp^{\rm short})^2}{12{r'}^3}.
\end{equation}
This difference in sign, then, implies that, one can arrange for the densities of far away and nearby bulk D-particles, which are not in general homogeneous, to be such that the total contribution to the brane world's vacuum energy is always subcritical, so that
 issues such as overclosure
of the universe by a significant population of D-particle defects
can be avoided.

For our purposes in this work we may therefore consider
that statistically significant populations of D-particles existed
in the early eras of the brane universe. As the time elapses, the
brane universe, which propagates in the higher-dimensional bulk (\emph{cf.}
fig.~\ref{fig:recoil}), enters regions characterised by D-particle
depletion, in such as way that the late eras cosmology of the universe
is not affected. Nevertheless, as we shall discuss below, the early
D-particle populations may still have important effects in generating
neutrino-antineutrino populations differences (asymmetries), which
are then communicated to the baryon sector via the standard sphaleron
processes \cite{PhysRevD.36.581}  or B-L conserving GUT symmetries in unified particle physics
models.

To this end, we need to consider the \emph{effective dispersion relation}
of a (anti)neutrino field in a brane space-time punctured with statistically
significant populations of D-particles. The latter is a dynamical
population, consisting of defects crossing the brane all the time,
thereby appearing to a brane observer as flashing ``on'' and ``off''
space-time ``foamy'' structures. The (anti)neutrino excitations
are represented as matter open strings with their ends attached on
the brane.The number density of (anti) neutrinos on the brane world
is limited by the requirement that they do not overclose the universe.
if neutrinos are assumed to have a chemical potential $\mu$, then
standard cosmological neutrino models predict that the number densities
of a single flavour of relativistic neutrinos \textbackslash{}emph\{plus\}
antineutrinos in thermal equilibrium at temperature $T_{\nu}$ is
estimated by \cite{Raffelt:2002nz}
\begin{equation}
n_{\nu\overline{\nu}}=T_{\nu}^{3}\frac{3\zeta_{3}}{2\pi^{2}}\Big(1+\frac{2\,{\rm ln}2\,\mu_{\nu}^{2}}{3T_{\nu}^{2}\zeta_{3}}+\frac{\mu_{\nu}^{4}}{72T_{\nu}^{4}\zeta_{3}}+\mathcal{O}\big(\frac{\mu_{\nu}^{6}}{T_{\nu}^{6}}\big)\Big)
\end{equation}
upon making the standard assumption that $\mu_{\nu}\ll T_{\nu}$
for all neutrino flavours. The quantity $\xi_{\nu}\equiv\frac{\mu_{\nu}}{T_{\nu}}$
is called the degeneracy parameter and is invariant under cosmic expansion.
if we assume that the electron-neutrino chemical potential is the
only one with significant presence in the early universe, then Big-Bang-Nucleosynthesis
(BBN) constraints imply $-0.04<\xi_{\nu_{e}}<0.07$. Thus, the order
of magnitude of the neutrino plus antineutrino number density is agrees
with naive standard estimate $n_{\nu\overline{\nu}}\sim\frac{3}{11}\, n_{\gamma}$,
where $n_{\gamma}$ is the photon density. Thus, today, where the
temperature of the universe is of order $T_{0}=2.728$\textasciitilde{}K
(Cosmic Microwave Background), corresponding to an energy of $k_{B}T_{0}\sim2.35\times10^{-13}~{\rm GeV}$
(with $k_{B}$ Boltzmann constant), the density of neutrinos is found
to be of order $n_{\nu{\overline{\nu}}}^{(0)}\sim112~{\rm cm}^{-3}$
and scales roughly with the cubic power of the temperature: $n_{\nu\overline{\nu}}\sim n_{\nu\overline{\nu}}^{(0)}\,\Big(\frac{T_{\nu}}{T_{0}}\Big)^{3}$.
So, for the decoupling temperatures of neutrinos, $k_{B}T_{d}\sim10^{15}$\textasciitilde{}GeV,
where we are interested in this work, in order to compute the frozen
CPT Violating neutrino-antineutrino population differences, one obtains
a number density of neutrino plus antineutrino populations of order
\begin{equation}
n_{\nu\overline{\nu}}(T=T_{d}\sim10^{15}\:\mathrm{GeV})\sim10^{85}\, cm^{-3}\,.\label{nupop}
\end{equation}
On the other hand, as already mentioned, there are no similar restrictions
on the population of the D-particle defects on the brane, in view
of the negative contributions on the potential energy of the brane
universe by bulk D-particle populations \cite{Mavromatos:2010nk}. Thus, at the early universe,
at the above neutrino-decoupling temperatures, we may even assume
D-particle densities of \textit{one} defect per \textit{string} volume
on the three brane world, without overclosing the universe. The assumption
that the string length can take on values in the phenomenologically
acceptable (post LHC era) range $10^{-27}-10^{-32}\,$cm, corresponding
to string mass scales from 10 TeV to 10$^{18}$ GeV, yields then a
D-particle number density in the range
\begin{equation}
n_{{\rm D}}(T=T_{d}\sim10^{15}~{\rm GeV})\sim10^{54}-10^{96}~{\rm cm}^{-3}\label{dpop}
\end{equation}
respectively. Thus we observe that in order to be able to treat the
D-particle populations as providing a more-or-less uniform ``medium''
over which neutrinos propagate, with non-trivial effective dispersion
relations, we need to have at the decoupling temperature much higher
densities of D-particles than those of neutrinos plus antineutrinos.
Comparing (\ref{nupop}) with (\ref{dpop}), we observe that, if one
assumes \textit{one} D-particle per \textit{three-dimensional string
volume} on the brane, then this latter requirement excludes the low
values of the string mass scale, implying an allowed range
\begin{equation}
10^{-5}M_{P}\le M_{s}\le10^{-1}M_{P}~,\label{msrange}
\end{equation}
 with $M_{P}\sim10^{19}\,$GeV the four-dimensional Planck mass. One
of course could have much more dense D-particle gases in the early
universe, which would allow for lower string scales.

\subsection{Kinematics of D-particle scattering and CPTV induced Leptogenesis}\label{sec:kinD}

We will now estimate the \emph{modification
of the dispersion relations} of neutrinos in such a ``media'' of D-particles
in the early universe. The interaction of a string with a D-particle implies that
at least one of the ends of the string  is attached to the D-particle defect. Furthermore,
the simultaneous creation of virtual strings stretched between the
defect and the brane,  describes the recoil of the D-particle.
During the interaction time, the D-particle undergoes motion characterized
by non-trivial velocities, $u_{\parallel}=\frac{g_{s}}{M_{s}}\Delta p_{i}=\frac{g_{s}}{M_{s}}r_{i}\, p_{i}$
along the brane longitudinal dimensions, where $r_{i}$ denotes the
proportion of the incident neutrino momentum that corresponds to the
momentum transfer $\Delta p_{i}$ during the scattering, and $v_{\perp}$
in directions transverse to the brane world \cite{Mavromatos:2005bu}  .

As discussed in ~\cite{Ellis:1999uh,Bernabeu:2006av,Mavromatos:2009rf} the non-trivial
capture and splitting of the open string during its interaction with
the D-particle, and the recoil of the latter, result in a \emph{local}
effective metric distortion of the form:
\begin{equation}
ds^{2}=g_{\mu\nu}dx^{\mu}dx^{\nu}=(\eta_{\mu\nu}+h_{\mu\nu})dx^{\mu}dx^{\nu}~,\qquad h_{0i}=(u_{i\,\parallel}^{a}\sigma_{a})~,\label{recmetric}
\end{equation}
where $u_{i\,\parallel}$ is the recoil velocity of the D-particle
\textit{on} the D-brane world, with $i=1,2,3$ a spatial space-time
index, $\sigma_{a}$ are the $2\times2$ Pauli flavour matrices with
$a=1,2,3$ (assuming two-flavour oscillations for simplicity). On
average over a population of stochastically fluctuating D-particles
including flavour changes, one may have the conditions (\ref{LI}),
the second of which in the case of flavour oscillations can be generalised
to
\begin{equation}
\ll u_{a,i}^{\parallel}u_{b,j}^{\parallel}\gg=\sigma^{2}\delta_{ij}\delta_{ab}~.\label{LI2}
\end{equation}
(We still assume that $\ll u_{a,i}^{\parallel}\gg=0$ . ) As a result
of (\ref{LI2}), on average, the flavour change during the interactions
of neutrinos with the D-foam can be ignored. In such a case, any flavour
structure in the metric (\ref{recmetric}) is ignored~\footnote{Ignoring the flavour structure, the metric (\ref{recmetric}) can be written as
\begin{equation}\label{waterfall}
ds^2 = dt^2 + 2 u_i dx^i dt - \delta_{ij} dx^i dx^j ~.
\end{equation}
This metric was determined from world-sheet conformal field theory considerations~\cite{Ellis:1999uh} and represents a dragging of the frame by the Gallilean (slowly moving) D-particle,
which moves on a flat space-time background. However, the string excitations represent relativistic particles, and as such they move according to the rules of special relativity.
Any four vectors attached to the strings, such as a four velocity, will evolve by a series of infinitesimal Lorentz boosts induced by the change of the D-particle velocity relative to the particle.
In this sense, one may perform a time coordinate change in the metric (\ref{waterfall}) to write in in the form, up to terms $u^3$ for small recoil velocities $|\vec u | \ll 1$:
\begin{equation}\label{waterfall2}
ds^2 = dt_{\rm ff} ^2 + 2 u_i dx^i dt_{\rm ff} - \delta_{ij} (dx^i - u^i dt_{\rm ff} ) (dx^j - u^j dt_{\rm ff}) + \mathcal{O}(u^3).
\end{equation}
The metric (\ref{waterfall2}) is nothing but the so-called Gullstrand-Painlev\'e metric~\cite{gpm}, representing the geometry in the exterior of a Schwarzschild black hole, where the
falling space into the black hole is represented as a Gallilean river on a flat space-time in which relativistic fishes swim. The river represents the frame of the recoiling D-particle, while the fishes are the relativistic matter strings. Here $t_{\rm ff}$ is the time of a free-floating observer who is at rest at infinity (compared to the centre of the black hole). In the case of a black hole the relative velocities $u^i$ are coordinate dependent, of course, unlike our approximation in the D-foam case, although one may easily consider more general cases, where the recoil velocities of the D-particles in the foam are non uniform, in which
case the analogy with the Gullstrand-Painlev\'e river would become stronger. }. This result is the motivation for the consideration of
a more general structure: Finsler geometry with stochastic parameters examined previously.

However, the effects of D-foam go beyond those encoded in the induced Finsler like metric.
The fine tuning that is required in stochastic Finsler metrics to get the correct
sign for the particle-antiparticle asymmetry is a feature, although commonplace
in other approaches, that is not entirely satisfactory.  However because
we have a microscopic model we can consider the kinematics of D-particle scattering.
On considering string theory scattering amplitudes we find that the four momentum is conserved
in the scattering of D-particles and strings. D-particles in the bulk exert forces on the vacuum energy
of the brane world of mixed sign, depending on their relative distance.
Thus, during the scattering process of a neutrino field with a D-particle,
the vacuum energy of the brane fluctuates by an amount $\Delta\mathcal{V}$
which depending on the process can be of either sign. From energy-momentum
conservation, at each individual scattering event between a neutrino
field and a recoiling D-particle, one could thus write:
\begin{equation}
\vec{p_{{\rm before}}}+\vec{p}_{{\rm after}}+\frac{M_{s}}{g_{s}}\,\vec{u}_{\parallel}=0~,\quad E_{{\rm before}}=E_{{\rm after}}+\frac{1}{2}\,\frac{M_{s}}{g_{s}}\,{\vec{u}}_{\parallel}^{2}+\Delta\mathcal{V}\label{neutrino}
\end{equation}
 where $(\vec{p},\, E)_{{\rm before\,(after)}}$ denote the incident
(outgoing) neutrino momenta, energies repectively and we used the
fact that the recoiling heavy D-particle of mass $M_{s}/g_{s}$ (with
$M_{s}$ the string scale and $g_{s}<1$ the string coupling, assumed
weak, so that string perturbation theory applies) has a non-relativistic
kinetic energy $\frac{1}{2}\,\frac{M_{s}}{g_{s}}\,{\vec{u}}_{\parallel}^{2}$.
We have also assumed that the fraction of the neutrino momentum transfer
in the direction perpendicular to the brane world is negligible. The
importance of the term $\Delta\mathcal{V}$ not having a fixed sign
in each individual scattering process is associated with the possibility
of D-particle induced neutrino flavour oscillations~\cite{Bernabeu:2006av}.

Indeed, upon averaging $\langle\langle\dots\rangle\rangle$ over a
statistically significant number of events, due to multiple scatterings
in a D-foam background, we may use the following stochastic hypothesis~\cite{Bernabeu:2006av}
\begin{equation}
\ll u_{i\,\parallel}\gg=0~,\qquad\ll u_{i\,\parallel}u_{j\,\parallel}\gg=\sigma^{2}\delta_{ij}~.\label{LI}
\end{equation}
 implying that Lorentz invariance holds only as an average symmetry
over large populations of D-particles in the foam. At a microscopic
level, (\ref{LI}) translates to momentum conservation on average
in (\ref{neutrino}), since $\ll\vec{u}_{\parallel}\gg=0$. At an
individual scattering process, if one represents the energy of the
incident neutrino on-shell as $\sqrt{{\overline{p}}^{2}+m_{1}^{2}}$,
where $\overline{p}$ is the amplitude of the conserved spatial momentum
of the neutrino, and the outgoing one as $\sqrt{{\overline{p}}^{2}+m_{2}^{2}}$,
we observe that the energy-conservation equation (\ref{neutrino})
implies in general $m_{1}\ne m_{2}$. Which one is larger depends
on the signature of the term $\frac{1}{2}\,\frac{M_{s}}{g_{s}}\,{\vec{u}}_{\parallel}^{2}+\Delta\mathcal{V}$,
which as mentioned is not of fixed sign, thereby allowing for neutrino
oscillations to take place. The situation is somewhat analogous to
the standard Mossbauer effect~\cite{Mossbauer:58}, where the emitted
or absorbed photon from a nucleus of an atom bound in a solid may
sometimes be free of nuclear recoil, in contrast to the case of gases,
thereby attributing the phenomena of nuclear resonances to such recoil-free
fraction of nuclear events. In our case the r�le of the ``nuclei
bound in a lattice'' is played by the D-particle lattice. In addition
to the D-particle recoil energy during scattering with stringy matter,
which would lead to energy losses for the neutrinos, there are vacuum
energy fluctuations, as a consequence of the motion of bulk particles
in the foam, thus the neutrino experiences losses and gains from the
vacuum, which results in the induced flavour oscillations. The analogue
of resonances in this case would correspond to the loss-and-gain-free
fraction of events, in which the neutrino does not oscillate.

However, the effects of the D-foam go beyond the above-mentioned kinematical
ones. On assuming
isotropic momentum transfer, $r_{i}=r$ for all $i=1,2,3$. The dispersion
relation of a neutrino of mass $m$ propagating on such a deformed
isotropic space-time, then, reads:
\begin{equation}
p^{\mu}p^{\nu}g_{\mu\nu}=p^{\mu}p^{\nu}(\eta_{\mu\nu}+h_{\mu\nu})=-m^{2}\,\Rightarrow\, E^{2}-2E{\vec{p}}\cdot u_{\parallel}-{\vec{p}}^{2}-m^{2}=0~.\label{dispersion}
\end{equation}
This on-shell condition implies that
\begin{equation}
E=\vec{u_{\parallel}}\cdot{\vec{p}}\pm\sqrt{({\vec{u}}\cdot{\vec{p}})^{2}+{\vec{p}}^{2}+m^{2}}~.\label{dispersion2}
\end{equation}
We take the average $\ll\dots\gg$ over D-particle populations with
the stochastic processes (\ref{LI2}), (\ref{LI}). Hence we arrive
at the following expression for an average neutrino energy in the
D-foam background:
\begin{eqnarray}
\ll E\gg & = & \ll\vec{p}\cdot{\vec{u}}\gg\pm\ll\sqrt{p^{2}+m^{2}+({\vec{p}}\cdot{\vec{u}})^{2}}\gg\nonumber \\
 & \simeq & \pm\sqrt{p^{2}+m^{2}}\left(1+\frac{1}{2}\sigma^{2}\right),\qquad p\gg m~,\label{dispersion3}
\end{eqnarray}
for the active light neutrino species. The last relation in eq.~(\ref{dispersion3})
expresses the corrections due to the space-time distortion of the
stochastic foam to the free neutrino propagation. It is this expression
for the neutrino energies that should be used in the averaged energy-momentum
conservation equation (\ref{neutrino}) that characterises a scattering
event between a neutrino and a D-particle. On further making the assumption
for the brane vacuum energy that $\ll\Delta\mathcal{V}\gg=0$,
the total combined effect on the energy-momentum dispersion relations,
from both capture/splitting and metric distortion, can then be represented
as:
\begin{equation}
\ll E_{2}\gg=\pm\sqrt{p^{2}+m^{2}}\left(1+\frac{1}{2}\sigma^{2}\right)-\frac{1}{2}\frac{M_{s}}{g_{s}}\,\sigma^{2}
\end{equation}
Since antiparticles of spin 1/2 fermions can be viewed as ``holes''
with negative energies, we obtain from (\ref{neutrino}) and (\ref{dispersion3})
the following dispersion relations between particles and antiparticles
in this geometry (for Majorana neutrinos, the r�les of particles /antiparticles
are replaced by left/right handed fermions):
\begin{eqnarray}
\ll E_{\nu}\gg & = & \sqrt{p^{2}+m_{\nu}^{2}}\left(1+\frac{1}{2}\sigma^{2}\right)-\frac{1}{2}\frac{M_{s}}{g_{s}}\,\sigma^{2}\nonumber \\
\ll E_{\overline{\nu}}\gg & = & \sqrt{p^{2}+m_{\nu}^{2}}\left(1+\frac{1}{2}\sigma^{2}\right)+\frac{1}{2}\frac{M_{s}}{g_{s}}\,\sigma^{2}\label{cptvdisp}
\end{eqnarray}
 where ${\overline{E}}>0$ represents the positive energy of a physical
antiparticle. In our analysis above we have made the symmetric assumption
that the recoil-velocities fluctuation strengths are the same between
particle and antiparticle sectors. (Scenarios for which this symmetry
was not asssumed have also been considered in an early work~\cite{Bernabeu:2006av}.)
There can thus be \emph{local} CPTV in the sense that the effective
dispersion relation between neutrinos and antineutrinos are different.
This is a consequence of the local violation of Lorentz symmetry (LV),
as a result of the non-trivial recoil velocities of the D-particle,
leading to the LV space-time distortions (\ref{recmetric}).

The discussion of CPTV in such foamy universes now follows the line
of argument adopted by others: the difference in the dispersion relations
between particles and antiparticles will imply differences in the
relevant populations of neutrinos ($n$) and antineutrinos (${\overline{n}}$), (\emph{cf.} the dispersion (\ref{cptvdisp})). This difference between neutrino
and antineutrino phase-space distribution functions in D-foam backgrounds
generates a matter-antimatter lepton asymmetry in the relevant densities
\begin{equation}
\ll n-{\overline{n}}\gg=g_{d.o.f.}\int\frac{d^{3}p}{(2\pi)^{3}}\ll[f(E)-f(\overline{E})]\gg~,\label{Leptonasym}
\end{equation}
 where $g_{d.o.f.}$ denotes the number of degrees of freedom of relativistic
neutrinos, and $\ll\dots\gg$ denotes an average over suitable populations
of stochastically fluctuating D-particles (\ref{LI}).

Let us first make the plausible assumption that $\sigma^{2}$ is constant
i.e. independent of space. It is a parameter which can only be positive.
Furthermore we will ssume that $\sigma^{2}$ is independent of the
(anti)neutrino energy. This is for estimation purposes only. We shall
come back to a more detailed analysis later. We should note that the
form of the dispersion relations (\ref{cptvdisp}) is analogous to
the case of CPTV axisymmetric geometries in the early universe, discussed
previously (\emph{cf.} (\ref{nunubardr})) with the r�le of the axial
curvature scalar potential $B_{0}$ being played here by the quantity
$\frac{1}{2}\frac{M_{s}}{g_{s}}\sigma^{2}$. In fact, it is easily
seen that to leading order in $\sigma^{2}$ the $(1+\frac{1}{2}\sigma^{2})$
prefactors of the square roots in (\ref{cptvdisp}) play no r�le,
and hence the leading in $\sigma^{2}$ contribution to the Leptonic
asymmetry comes from the constant $\frac{1}{2}\frac{M_{s}}{g_{s}}\,\sigma^{2}$
terms in the dispersion relations. The induced lepton asymmetry can
therefore be calculated following similar steps as those leading to
(\ref{dlbianchi}), upon the replacement of $B_{0}$ by $\frac{1}{2}\frac{M_{s}}{g_{s}}\sigma^{2}$,
the difference being that here the value of $B_{0}$ (\emph{i.e.}
of the D-foam recoil fluctuations $\sigma^{2}$) is to be fixed phenomenologically.

The result for the D-foam-induced lepton asymmetry can be estimated
from (\ref{Leptonasym}), using (\ref{cptvdisp}). Ignoring neutrino
mass terms and $(1+\frac{\sigma^{2}}{2})$ square-root prefactors
in (\ref{cptvdisp}), setting the (anti)neutrino chemical potential
to zero (which is a sufficient approximation for relativistic light
neutrino matter) and performing a change of variables $|\vec{p}|/T\to\tilde{u}$
we obtain from (\ref{Leptonasym}) the result:
\begin{eqnarray}\label{dn2}
\Delta n_{\nu} & = & \frac{g_{d.o.f.}}{2\pi^{2}}\, T^{3}\int_{0}^{\infty}d\tilde{u}\,\tilde{u}^{2}\,[\frac{1}{1+e^{\tilde{u}-\frac{M_{s}\sigma^{2}}{2g_{s}\, T}}}-\frac{1}{1+e^{\tilde{u}+\frac{M_{s}\sigma^{2}}{2\, g_{s}\, T}}}]=\frac{g_{d.o.f.}}{\pi^{2}}\, T^{3}\left({\rm Li}_{3}(-e^{-\frac{M_{s}\sigma^{2}}{2g_{s}\, T}})-{\rm Li}_{3}(-e^{\frac{M_{s}\sigma^{2}}{2g_{s}\, T}})\right)\nonumber \\
 & \simeq & \frac{g_{d.o.f.}}{\pi^{2}}\, T^{3}\left(\frac{M_{s}\sigma^{2}}{g_{s}\, T}\right)\,>\,0,
\end{eqnarray}
 to leading order in $\sigma^{2}$, where in the last step we took
into account the formal definition as a series of the Polylogarithm
function ${\rm Li}_{s}(z)=\sum_{k=1}^{\infty}\,\frac{z^{k}}{k^{s}}$
which is valid for $|z|\,<\,1$, while the cases $|z|\ge1$ are defined
by analytic continuation. We thus observe that the CPTV term $-\frac{1}{2}\frac{M_{s}}{g_{s}}\sigma^{2}$
in the dispersion relation (\ref{cptvdisp}) for the neutrino, which
corresponds to the energy `loss' due to the D-particle recoil kinetic
energies, comes with the right sign (`loss') so as to guarantee an
excess of particles over antiparticles. Unlike the model of ~\cite{Debnath:2005wk,Mukhopadhyay:2005gb,Mukhopadhyay:2007vca,Sinha:2007uh},
then, where the sign of the $B_{0}$ parameter had to be assumed,
in our D-foam case there is no such freedom, and the positive $\Delta n_{\nu}$
is derived from first principles. We consider this a nice feature
of our model.

As in standard scenarios of Leptogenesis, the Lepton asymmetry (\ref{dn2})
decreases with decreasing temperature up to a freeze-out point, which
occurs at temperatures $T_{d}$ at which the Lepton-number violating
processes decouple. This is taken to be the conventional one (in standard
scenarios of Leptogenesis): $T_{d}\sim10^{15}$~GeV.

The resulting lepton asymmetry then freezes out to a value (\emph{cf}.
Eq.~(\ref{dlbianchi}) ):
\begin{equation}
\Delta L(T<T_{d})=\frac{\Delta n_{\nu}}{s}\sim\frac{M_{s}\,\sigma^{2}}{g_{s}\, T_{d}}\label{Ddl}
\end{equation}
 which survives today. The so calculated $\Delta L$ assumes the phenomenologically
relevant order of magnitude of $10^{-10}$ provided
\begin{equation}
\frac{M_{s}}{g_{s}}\,\sigma^{2}\sim10^{5}\,{\rm GeV}~.\label{conds}
\end{equation}
 Since in these scenarios, the dimensionless stochastic variable,
expressing fluctuations of a recoil velocity, is always less than
one, $\sigma^{2}<1$, one observes that the required Lepton asymmetry
is obtained for D-particle masses larger than
\begin{equation}
\frac{M_{s}}{g_{s}}\,>\,100~{\rm TeV}~.
\end{equation}
 The so obtained $\Delta L$ can then be communicated to the baryon
sector, t yield the observed (today) baryon asymmetry, via either
B+L violating sphaleron processes, or B-L conserving interactions
in Grand Unified theories, as in standard scenarios.

The above estimate ignores an important fact, namely the dependence
of the stochastic variable $\sigma^{2}$ on the neutrino energy. Indeed,
as discussed in detail in \cite{Mavromatos:2005bu,Bernabeu:2006av},
one may parameterise the momentum transfer by the fraction parameter
of the incident momentum $r$, which is in turn assumed stochastic,
that is
\begin{equation}
u_{i}=\frac{g_{s}}{M_{s}}\Delta p_{i}\rightarrow g_{s}\, r_{i}\frac{p_{i}}{M_{s}}~,{\rm no~sum~over~i}~,\quad\ll r_{i}\gg=0~,\quad\ll r_{i}r_{j}\gg=
{\Delta}^{2}\delta_{ij}~.\label{rstoch}
\end{equation}

In this case, the dispersion relations (\ref{cptvdisp}) are modified
by the replacement of
\begin{equation}\label{dsigma}
\sigma^{2}\rightarrow\frac{g_{s}^{2}}{M_{s}^{2}}\,{\Delta}^{2}p^{2}~,
\end{equation}
which is now momentum dependent:
\begin{eqnarray}\label{cptvdisp2}
\ll E_{\nu}\gg & = & \sqrt{p^{2}+m_{\nu}^{2}}\left(1+\frac{g_{s}^{2}}{2\, M_{s}^{2}}{\Delta}^{2}\, p^{2}\right)-\frac{g_{s}}{2\, M_{s}}{\Delta}^{2}\, p^{2}\nonumber \\
\ll E_{\overline{\nu}}\gg & = & \sqrt{p^{2}+m_{\nu}^{2}}\left(1+\frac{g_{s}^{2}}{2\, M_{s}^{2}}{\Delta}^{2}\, p^{2}\right)+\frac{g_{s}}{2\, M_{s}}{\Delta}^{2}\, p^{2}
\end{eqnarray}
 Below, we shall evaluate the integral (\ref{Leptonasym}) for the
case (\ref{LI}), assuming again ${\Delta}^{2}\ll1$ sufficiently
small so that a truncation to order ${\Delta}^{2}$ will
be sufficient. ${\Delta}^{2}$ will be assumed to be the
same between particle and antiparticle sectors.

In contrast to conventional point-like field theory models, where
the upper limit of momentum integration can be extended to $\infty$,
in D-foam models, due to (\ref{recmetric}), this is extended up to
the value for which the D-particle recoil velocity approaches the
value of the speed of light\emph{ in vacuo}, c=1 in our units, \emph{i.e.}
\begin{equation}
p_{{\rm max}}\equiv|\vec{p}|_{{\rm max}}=\frac{M_{s}}{g_{s}\,\sqrt{{{\Delta}}^{2}}}~,
\end{equation}
 where $r$ is the stochastic variable satisfying (\ref{rstoch}).
The resulting integrals in (\ref{Leptonasym}) then become:
\begin{equation}
\Delta n_{\nu}=\frac{g_{d.o.f.}}{2\pi^{2}}\, T^{3}\int_{0}^{\frac{M_{s}}{T\, g_{s}\,\sqrt{{\Delta}^{2}}}}\, d\tilde{u}\,\left(\frac{1}{1+e^{\tilde{u}-\tilde{u}^{2}\,\frac{g_{s}{\Delta}^{2}T}{2M_{s}}}}-\frac{1}{1+e^{\tilde{u}+\tilde{u}^{2}\,\frac{g_{s}{\Delta}^{2}T}{2M_{s}}}}\right)\simeq....\label{dn3}
\end{equation}
 The upper limit of the $\tilde{u}$ integration cannot be taken as
$\infty$ since $\Delta n_{\nu}$ should be evaluated at
the decoupling temperature $T_{d}\sim10^{15}~{\rm GeV}$, where it
freezes out, as mentioned previously. An advantage of this CPTV as
compared to the ones associated with axisymmetric geometries in the
early universe~\cite{Debnath:2005wk,Mukhopadhyay:2005gb,Mukhopadhyay:2007vca,Sinha:2007uh}
is that the CPTV parameter, ${\Delta}^{2}$, which characterises
the fluctuations in the neutrino momentum transfer due to its interactions
with the D-foam, is a parameter rather than a function. It is phenomenological
(since it depends on the density of D-particles in the early universe)
and cannot be significantly constrained by the cosmology of the early
universe. We shall now analyse (\ref{dn3}); first we set
\[
y\equiv\frac{M_{s}}{T\, g_{s}\,\sqrt{{\Delta}^{2}}}\left(\equiv\frac{\alpha}{\sqrt{{\Delta}^{2}}}\right).
\]
%and then write
%\begin{equation}
%\mathcal{I}\left(y,\xi\right)\equiv\int_{0}^{y}\frac{1}{1+\exp\left(u-\frac{\xi}{2y}u^{2}\right)}du.\label{dn4}
%\end{equation}
We note that
\begin{eqnarray}
  \Delta  n_{\nu} &=& \frac{g_{d.o.f.}}{2\pi^{2}\,} T^{3}\left[\int_{0}^{y}\frac{du}{1+e^{u-u^{2}\Delta/{2y}  }}-\int_{0}^{y}\frac{du}{1+e^{u+u^{2}\Delta/{2y}  }}\right], \label{dn5} \\
   &=& \frac{g_{d.o.f.}}{2\pi^{2} \,} T^{3} y\left[\int_{0}^{1}\frac{dv}{1+e^{\frac{\alpha v}{\Delta}-\frac{\alpha v^{2}}{2}}} - \int_{0}^{1}\frac{dv}{1+e^{\frac{\alpha v}{\Delta}+\frac{\alpha v^{2}}{2}} }  \right]
\end{eqnarray}
%\begin{equation}
%\Delta  n_{\nu}=\frac{g_{d.o.f.}}{2\pi^{2}\, T^{3}}\left[\int_{0}^{y}\frac{du}{1+e^{u-u^{2}\Delta/2y  }}}-\int_{0}^{y}\frac{du}{1+e^{u+u^{2}\Delta/2y  }}}\right].\label{dn5}
%\end{equation}
We write $\mathcal{I}_{1}= \int_{0}^{1}\frac{dv}{1+\exp(\frac{\alpha v}{\Delta}-\frac{\alpha v^{2}}{2})}$ and $\mathcal{I}_{2}=\int_{0}^{1}\frac{dv}{1+\exp(\frac{\alpha v}{\Delta}+\frac{\alpha v^{2}}{2}) } $. We will evaluate $\Delta  n_{\nu}$ in two limits:(i) $\alpha \rightarrow \infty, \; \Delta \sim O(1)$ and (ii) $\Delta \rightarrow 0, \; \frac{\Delta}{\alpha}\rightarrow 0$.
%where
%$\mathcal{J}\left(y,{\Delta}\right)=\int_{0}^{1}\left(1+\exp\left(y\left[v-v^{2}{\Delta}/2\right]\right)\right)^{-1}dv$.
%$Now as ${\Delta}\rightarrow0$ from Watson's lemma
%\[
%\mathcal{J}\left(y,{\Delta}\right)\sim\sum_{n=0}^{\infty}\frac{\Delta^{2n+1}}{2^{n}n!}\frac{\Gamma\left(2n+1\right)}{\alpha^{n+1}}
%\]
%and so
%\[
%\mathsf{\mathbb{\mathrm{\Delta}}}n_{\nu}\sim\frac{g_{d.o.f.}}{\pi^{2}}\, %T^{3}\sum_{n=0}^{\infty}\frac{\Delta^{2n}}{2^{n}n!}\frac{\Gamma\left(2n+1\right)}{\alpha^{n}}.
%\]
For case (i) let $x= \frac{v}{\Delta} -\frac{v^{2}}{2}$ and so
\begin{equation}
    \mathcal{I}_{1}=\int_{0}^{\frac{1}{\Delta}-\frac{1}{2}}\frac{dx}{(\frac{1}{\Delta}-v)}\frac{1}{1+\exp(\alpha x)}
\end{equation}
where $v$ is given by
\[
v(x)=\frac{1-\sqrt{1-2x \Delta^{2}}}{\Delta}.
\]
This integral can be evaluated (using Watson's lemma) to give
\begin{equation}\label{I1}
    \mathcal{I}_{1}\sim \Delta (\frac{1}{\alpha}+\frac{\Delta^{2}}{\alpha^{2}}+\frac{3 \Delta^{4}}{\alpha^{3}}+  \ldots ).
\end{equation}
A similar analyis for $\mathcal{I}_{2}$ gives
\begin{equation}\label{I2}
    \mathcal{I}_{2}\sim \Delta (\frac{1}{\alpha}-\frac{\Delta^{2}}{\alpha^{2}}+\frac{3 \Delta^{4}}{\alpha^{3}}+  \ldots ).
\end{equation}
Hence $\Delta  n_{\nu}$ in limit (i) is given by
\begin{equation}\label{deltaN1}
    \Delta  n_{\nu}=\frac{g_{d.o.f.}}{\pi^{2}\,} T^{3}\frac{\Delta^{2}}{\alpha}+ \ldots =
    \frac{g_{d.o.f.}}{\pi^{2}\,} T^{3}\frac{\Delta^{2}\, g_s T}{M_s}
\end{equation}
For case (ii) we have
\begin{equation}\label{I12}
    \mathcal{I}_{1}\sim \int_{0}^{1}e^{-\frac{\alpha}{\Delta}v}e^{\frac{\alpha v^{2}}{2}}\qquad {\rm and} \qquad
    \mathcal{I}_{2}\sim \int_{0}^{1}e^{-\frac{\alpha}{\Delta}v}e^{-\frac{\alpha v^{2}}{2}}.
\end{equation}
However the analysis again leads to the result (\ref{deltaN1}). The lepton asymmetry resulting from (\ref{deltaN1}) freezes out  at temperature $T_{d}$ is
\begin{equation}\label{dlf}
    \Delta L(T<T_{d})=\frac{\Delta n_\nu }{s} = \frac{2 \Delta^{2}g_{s}T_{d}}{M_{s}}.
\end{equation}
We note that this result is compatible formally with (\ref{Ddl}) if one takes into account (\ref{dsigma}) and associates
the momentum maqnitude $p$ of the realtivistic neutrino (\emph{i.e}. its energy) with the temperature $T$.

From (\ref{dlf}), we observe that for a freeze-out temperature $T_d \sim 10^{15}$~GeV, the phenomenological value $\Delta L \sim 10^{-10}$ is attained for
\begin{equation}\label{lgv}
\frac{M_s}{g_s} \sim  10^{25} \, \Delta ^2 \, {\rm GeV}~.
\end{equation}
For $\Delta^2 \sim 10^{-6} $ a Planck size D-particle mass $M_s/g_s \sim 10^{19}$~GeV is required so that the D-foam provides the physically observed
Lepton and, thus, Baryon Asymmetry. For the unnaturally small $\Delta ^2 < 10^{-21}$ one arrives at $M_s/g_s \sim 10 ~{\rm TeV}$.
Unfortunately, for $\Delta^2 \sim \mathcal{O}(1)$ transplanckian D-particle masses are required.
We should stress that the above conclusions were based on the assumption that the freeze-out temperature was the temperature at
decoupling of neutrinos in standard big-bang cosmology.

Our approach to leptogenesis  is distinguished from others in that a local effective field theoretical
description is not adopted. Because of D-particle recoil when scattering off matter strings the background of D-particles  can
be modelled as a stochastic medium \cite{Mavromatos:2005bu,Mavromatos:2010nk,Bernabeu:2006av}. The underlying string theoretic description provides the rigorous description of the scattering of D-particles.
The D-particles backreact (as seen from infra-red divergences in perturbation theory) and change the metric which influences the space in which matter is moving. Furthermore as discussed at length in \cite{Mavromatos:2010nk}, and mentioned briefly above, the D-particle foam model does \emph{not} lead to overclosing the universe. Hence despite having statistically significant populations of D-particles in the early universe, which provide the CPTV background on which neutrinos propagate, the assumption of a subcritical energy density for the universe  can still hold.

\section{Conclusions and Outlook \label{sec:concl} }

In this work we have considered some models leading to CPTV   gravitational leptogenesis, based on string-inspired constructions.
The primary model that we discuss here involves a brane universe propagating in a higher-dimensional bulk space-time, which contains populations of
(``point-like'' ) D0 brane defects (D-particles).  The topologically non-trivial  interaction of the D-particles with the neutrinos, in the sense of capture and subsequent re-emergence of the latter by the defects, is allowed by electric charge conservation; in contrast the medium of defects is "transparent" for charged excitations of the standard model (such as charged leptons and quarks). The propagation of neutrinos in this medium of space-time defects results in CPTV dispersion relations, which are different for neutrinos and antineutrinos. It is essential to notice that this difference is obtained by considering the kinematics of the defect/neutrino scattering on the assumption that the antiparticle has negative energies following the``hole theory'' of Dirac. These considerations apply only to fermions, as opposed to bosons, because the ``hole theory''  is based on the fermion exclusion principle. This model has two aspects. One is the induced metric in space-time, which depends on momenta as well as space-time coordinates, and the other is the particle-like kinematics involved in the D0-brane scattering of matter.
The first aspect is similar to stochastic Finsler metrics which we have considered in a general way without necessarily tying it to the brane  model.
The second, kinematical aspect is important in leading naturally to matter dominance over antimatter, in contrast to earlier proposals of gravitational leptogenesis, where this sign has to be adjusted by hand.

The cosmology and particle-physics
phenomenology of such scenarios , although currently in their infancy, are worth pursuing in our opinion. In particular, for the defect model, at early epochs of the universe, despite the fact that significant populations of massive D-particles are assumed present, we have argued here and in our previous works \cite{Mavromatos:2010nk,Mavromatos:2010jt} that one can avoid overclosure of the  brane universe: the defects are allowed to propagate in the bulk and thus they exert forces on the brane world which are such that there are mixed sign contributions to the brane vacuum energy, depending on the distance of the bulk defect populations from the brane. The population of D-particles in the bulk is also constrained by the amount of Lorentz violation at late eras of the universe, which leave their imprint on the Cosmic Microwave Background~\cite{Lim:2004js,Kahniashvili:2008va} or in vacuum-refractive-index tests of arrival times of cosmic photons~\cite{Albert:2007kw,Albert:2007qk,Ackermann:2010us,Mavromatos:2010pk}.

\section*{Acknowledgments}

The work of N.E.M. was supported in part by the London Centre for
Terauniverse Studies (LCTS), using funding from the European Research
Council via the Advanced Investigator Grant 267352. NEM and SS also thank the 
STFC UK for partial support under the research grant ST/J002798/1.

\addcontentsline{toc}{chapter}{{\bf Bibliography}}
\bibliography{MSCPTVUniv2}
\bibliographystyle{apsrev4-1} 

\end{document}